\documentclass[showpacs,amsmath,amssymb,nofootinbib]{revtex4-1}
\usepackage{amsmath} 
\usepackage{graphicx}
\usepackage{subfig}
\usepackage{float}
\def\be{\begin{equation}}
\def\ee{\end{equation}}
\def\bea{\begin{eqnarray}}
\def\eea{\end{eqnarray}}

\def\gnl{g_{\rm NL}}

\def\taunl{\tau_{\rm NL}}

\def\vp{\varphi}
\def\vs{\varphi_{s}}
\def\vl{\varphi_{L}}
\def\dl{\delta_{L}}

\def\fnl{f_{\rm NL}}

\def\gnl{g_{\rm NL}}

\def\taunl{\tau_{\rm NL}}

\usepackage{color}

\begin{document}

\title{Primordial non-Gaussianity in the 
bispectra of large-scale structure}

\author{Gianmassimo Tasinato, Matteo Tellarini, Ashley J. Ross, and David Wands}
\email[E-mails: ]{gianmassimo.tasinato@port.ac.uk, matteo.tellarini@port.ac.uk, ashley.ross@port.ac.uk, david.wands@port.ac.uk}
\bigskip
\vskip1cm

\affiliation{
\vskip0.3cm
Institute of Cosmology \& Gravitation, University of Portsmouth, Dennis Sciama Building, Portsmouth, PO1 3FX, United Kingdom
}


\begin{abstract}
The statistics of large-scale structure in the Universe can be used to probe non-Gaussianity of the primordial density field, complementary to existing constraints from the cosmic microwave background.
 In particular,
   the scale dependence of halo bias, which affects the halo distribution at large scales, represents a promising tool for analyzing primordial non-Gaussianity of local form. Future observations, for example, may be able to constrain the trispectrum parameter  $\gnl$ that is difficult to study and constrain using the CMB alone. We investigate how galaxy and matter bispectra can distinguish between the two non-Gaussian parameters $\fnl$ and 
$\gnl$, whose effects give nearly degenerate contributions to the power spectra.   We use a generalization of the univariate bias approach, making the hypothesis that the number density of halos forming at a given position is a function of the local matter density contrast and of its local higher-order statistics.  Using this approach, we calculate the halo-matter bispectra and analyze their properties. We determine a connection between the sign of the halo bispectrum on large scales and the parameter $\gnl$. We also construct a combination of halo and matter bispectra that is sensitive to $\fnl$, with little contamination from $\gnl$. We study both the case of single and multiple sources to the primordial gravitational potential, discussing how to extend the concept of stochastic halo bias  to the case of bispectra. We use a specific  halo mass-function to calculate numerically the bispectra in appropriate squeezed limits, confirming our theoretical findings.
%
 \end{abstract}


\maketitle
\section{Introduction}

The intrinsic non-Gaussianity of primordial curvature perturbations is
 a valuable tool to distinguish among different models for the origin of structure in the very early universe~\cite{Bartolo:2004if}.   
Even if the primordial curvature perturbation has a Gaussian distribution, the initial density perturbation is non-Gaussian due to the non-linearity of the Einstein field equations \cite{Verde:2009hy,Bartolo:2010rw,Bruni:2013qta}.
Non-Gaussianity (nG) may be characterized by a variety of different parameters, 
    which control the departure of
     the underlying probability distribution function (pdf) of  primordial fluctuations from a purely Gaussian 
    distribution. 
     These parameters can be related to the amplitude of connected 
 $n$-point ($n$-pt) correlation functions of  primordial  curvature fluctuations 
 that vanish for $n>2$ in the Gaussian case. The connected $n$-pt functions may be scale- and shape-dependent, in a manner determined
 by the particular model that generates them.  

In this work, we will focus on the so-called local shape of nG, in which the  primordial gravitational potential
   in coordinate space  
   can be expressed
  as an expansion in powers of one or more Gaussian random fields that determine the primordial density fluctuations.
   In this case 3-pt functions of curvature fluctuations are determined 
 by a parameter, $\fnl$, that at lowest order in a perturbative expansion characterizes the skewness of the
 pdf of primordial fluctuations. 
 4pt functions are determined by parameters, $\gnl$ and $\taunl$, 
 controlling the kurtosis of the pdf at lowest order in perturbations.  
The recent release of Planck data allows one to set the best constraints yet on primordial non-Gaussian parameters from cosmic microwave background (CMB) data alone \cite{Ade:2013ktc}.
The quantity $\fnl$ (local) is now constrained to be $ |\fnl|\le 15$ at 
2$\sigma$ level, while $\taunl\le 2800$;
 an analysis of the constraints on  $\gnl$ local from Planck data
 has yet to be completed, but the forecast
 $1\sigma$
 error
 bar with Planck data 
 has been
 estimated to be
 $\sigma_{\gnl}	= 6.7 \times 10^4$ by \cite{Sekiguchi:2013hza} and 
  $\sigma_{\gnl}	= 1.3 \times 10^5$
 by \cite{Smidt:2010ra}.
   It may be unlikely that we live in a universe with
 a large hierarchy  between $\fnl$ and $\gnl$ local \cite{Nurmi:2013xv}, but this possibility cannot be
  excluded {\it a priori} and there are theoretical models able to predict this pattern (see for 
  example \cite{Byrnes:2009qy}
  or
  the discussion in \cite{Suyama:2013nva}). 
  Since $\fnl$ and $\gnl$ control distinct features of the pdf (skewness and kurtosis) it is 
  very important to have the best observational constraints on each of  them. 
  Complementary observables have to be considered to set more stringent bounds on non-Gaussian
  parameters, in particular on $\gnl$, and one possibility is
  to use the statistics of large-scale structure. 
  
       
  Pioneering papers by Dalal et al \cite{Dalal:2007cu}, Slosar et al \cite{Slosar:2008hx}, Matarrese
  et al \cite{Matarrese:2008nc}, Afshordi and Tolley  \cite{Afshordi:2008ru}
  highlighted an interesting feature of primordial nG of local shape: 
  it
  introduces a 
  specific correlation between modes of different wavelengths which leads to a 
  characteristic scale dependence of the halo bias.
    Much work has been done so far to explore this interesting topic (see \cite{Gong:2011gx,Desjacques:2008vf,Desjacques:2011jb,Desjacques:2011mq} and \cite{Desjacques:2010jw,Verde:2010wp} for reviews).  However it turns out to be 
   challenging to disentangle the contributions from different non-Gaussian parameters
   using only halo and matter power spectra, since the
     characteristic scale dependence of the bias 
    is primarily sensitive to a particular combination of the
   $\fnl$ and $\gnl$ parameters    \cite{Desjacques:2009jb,Roth:2012yy}.  
     Mild  corrections associated with the red-shift dependence of the halo mass function
    have been used to set the first LSS constraints on $\gnl$ in \cite{Giannantonio:2013uqa}, but
    it is difficult to convincingly distinguish the effect of $\fnl$ from that of $\gnl$
    using only
    galaxy power spectra.
      Independent constraints on $\fnl$ from CMB
    are useful, but not conclusive because $\fnl$ could be characterized by a significant scale dependence \cite{LoVerde:2007ri,Byrnes:2010ft}, that makes
    its value  probed at LSS scales different with respect to the one tested at CMB scales. 
   
A promising method that may break the degeneracy between $\fnl$ and $\gnl$ 
     is to study the bispectra of halo and matter densities, which
   are sensitive to a non-linear bias parameter that depends specifically on $\gnl$, and allows us to break the aforementioned 
   degeneracy.   
Jeong and Komatsu \cite{Jeong:2009vd} (see also \cite{Sefusatti:2009qh}) were the first 
 among various  groups 
to include
the scale dependence of 
 halo bias
  when studying bispectra, 
 by
considering the non-linear evolution of the halo overdensity in a local, univariate bias expansion,
where the halo abundance is taken to be a function of the local density.
They have shown that
galaxy bispectra are sensitive to non-Gaussian parameters beyond $\fnl$, as confirmed by $N$-body simulations \cite{Nishimichi:2009fs,Sefusatti:2011gt}. 
 In this work, we elaborate on this subject. 
  We will show that 
    halo and matter bispectra 
   have interesting qualitative features that may allow us to
    distinguish between the effects of different 
    primordial nG parameters.
%

 We implement a peak-background split method
 within a barrier crossing approach to re-derive in part the
 results of  \cite{Jeong:2009vd}  in a physically transparent way, and to extend their analysis in various directions with the main aim of understanding how the parameters
 $\fnl$ and $\gnl$ can be possibly distinguished when analyzing the statistics of halo and matter bispectra.  

In order to focus only on  the consequences  of primordial non-Gaussian initial conditions on the properties
of bispectra, we will not include the effects of non-linear gravitational clustering in our analysis; that is, we work in Lagrangian space and linearly transform to Eulerian space. 
  We also neglect non-linearity in the halo mass function, i.e., non-linear local bias, which implies that any non-vanishing bispectra will be solely due to primordial nG.

Our work contains several results, and we present these in a modular way to render the paper 
 easier to read. We begin with some theoretical sections discussing how to implement the peak-background split and
barrier crossing approaches to determine concise analytical expressions for halo and matter bispectra, when the gravitational potential is a function of  either single or multiple sources (e.g., due to fluctuations of single or multiple fields during inflation in the very early universe). Our formulae are obtained by applying to bispectra
the methods developed in \cite{Baumann:2012bc} for power spectra.  
 See also \cite{Scoccimarro:2011pz} for an interesting paper discussing how  to implement the peak background
splitting approach to bispectra.

We use  a generalization
of the univariate bias approach, making the hypothesis that the local number of halos 
forming at a given position 
is a function of the matter contrast, {\it and} of its local correlation functions evaluated at that
position. 
This is an alternative to the bivariate approach \cite{Giannantonio:2009ak} where the halo abundance is assumed to be a function of both the density field and the gravitational potential.
In practice, the physical effect of long-wavelength modes of the potential is to modulate the small-scale variance of the density field in local models of nG.
Our extended univariate approach is physically transparent, and allows us to recover some of the results of the bivariate approach, in appropriate limits.  In addition we can include local variations of effective non-Gaussianity parameters on halo abundance \cite{Byrnes:2011ri}.
The resulting bispectra contain several contributions scaling with different
powers of the scale $k$, weighted by  coefficients depending on primordial non-Gaussian parameters.
 In the limit of large scales, in which analytic expressions for the transfer functions can be derived,
 we will discuss how bispectra have qualitative features 
   that make it possible, at least in principle, to distinguish the effects of $\fnl$ and $\gnl$.
For example, by studying the properties of the bispectra as a function of the scale, we determine a connection between
 the sign of the halo bispectrum and nG parameters, in particular the value of $\gnl$. Moreover, we show that there 
 exists a combination of halo and matter bispectra that is sensitive to $\fnl$ only, 
allowing us to probe $\fnl$ without contaminations from $\gnl$. 
  
We then extend all these results, obtained in the case of a single source for the 
primordial gravitational potential, to a two-field example that allows us to understand how multiple sources contribute to galaxy bispectra. 
In \cite{Tseliakhovich:2010kf,Baumann:2012bc}
  it was pointed out  that  if multiple fields source the primordial gravitational potential,   then
   halo overdensities need not be fully correlated with matter overdensities.
We show that a combination of halo and matter power spectra  is particularly convenient for studying the phenomenon of {\it stochastic halo bias}.     
Our analysis enable us to extend the concept of stochastic halo bias in the presence of primordial nG to the cases of both power spectra and bispectra, including the effect of $\gnl$. 
The fact that this stochastic bias is non-vanishing 
   can be interpreted in terms of inequalities satisfied by
  primordial non-Gaussian parameters, and these provide information on the number
  of fundamental fields sourcing nG. 
 We determine a new combination of halo and matter bispectra
  whose value can be used to distinguish the effect of single and multiple sources on the primordial gravitational potential,
  and discuss its physical interpretation.

  
  In the second part of our work, we apply our theoretical findings  to a systematic  numerical
    analysis  of bispectra in 
  appropriate squeezed limits, by using
     Press-Schechter
   halo mass functions supplemented by non-Gaussian corrections controlled by an Edgeworth expansion. 
    This discussion  enables us to apply our results to  scales  and
    redshifts 
     that can be probed in present or future surveys.  
     By plotting the resulting
   halo bispectra, we explore  the  different roles played by  $\fnl$ and $\gnl$.
  The numerical analysis
    confirms our analytical results. We show  that
     qualitative features of the 
      profiles of halo and matter bispectra as a function of the scale 
are sensitive to different nG parameters, possibly allowing us to test      
 them individually.
   These results clearly demonstrate how galaxy bispectra have  
       features that could distinguish between different non-Gaussian parameters, in a way that is not
       possible by studying only the scale dependence of galaxy bias in
        power spectra of halos and matter.
        
        
        We conclude with a discussion       
     on possible improvements and future directions for our investigation.

\section{The statistics of dark matter halos}
    
    We start with an introductory section outlining the ideas  
     that we will use to treat the statistics of dark matter halos \cite{Baumann:2012bc}. 
  The primordial gravitational potential $\Phi$ is related to the linearly evolved
  matter density contrast $\delta$ in terms of the Poisson equation in Fourier space 
  \be
  \delta_{\vec k}(z)\,=\,\bar{\alpha}(k,\,z)\,\Phi_{\vec k}\,,
  \ee 
  where 
  \be
  \bar{\alpha}(k, z)\,=\,\frac{2\,k^2\,T(k)\,D(z)}{3\,\Omega_m\,H_0^2}\,.
  \ee
   $T(k)$ 
is the matter transfer function normalized such that $T(k)\to1$ at  large scales $k\to0$. $D(z)$ is the linear growth factor, as a
function of redshift $z$. It is normalized such that $D(z)\,=\,(1+z)^{-1}$ in matter domination. At very large scales $k\ll1$, the function
$\bar{\alpha}$ is proportional to  $k^2$ \cite{Bernardeau:2001qr}:
\be\label{formulal}
\bar{\alpha}(k,z)
 \simeq\frac{2k^2D(z)}{3\Omega_mH_0^2}=\alpha_0(z)k^2 \quad \text{where} \quad\alpha_0(z)\simeq2.16\cdot10^7\,\left(\frac{0.277 D(z)}{\Omega_m}\right)\,\left(\frac{\text{Mpc}}{h}\right)^2\,.
\ee
 This asymptotic behavior will be useful to get analytic results in the discussion of Section \ref{subsec:squeezed}. 
 To render the formulae less cumbersome, 
 from now on we will suppress the $z$ dependence. 
 
 The field $\delta_M(\vec{x})$ denotes the linear density contrast smoothed with a top-hat filter of
radius $R_M\,=\,(3M/4\pi \bar{\rho}_m)^{1/3}$ where $M$
is the halo mass, and $\bar{\rho}_m$ the background matter density. In Fourier space, we write
\be
\delta_M(\vec{k})\,=\,W_M(k)\,\delta_{\vec{k}}\label{smrel}
\ee
 with $W_M$ the Fourier transform of top-hat filter:
 \be
 W_M(k)\,=\,3\,\frac{\sin{(k R_M)}  -k R_M\,\cos{(k R_M)}}{\left( k \,R_M\right)^3}
 \,.
 \ee 
  We also define $\sigma_M\equiv \langle \delta_M^2\rangle^\frac12$ and $\alpha(k)\,=\,W_M(k) \bar{\alpha}(k)$.  
   At large scales $W_M(k\to0)\,=\,1$. 
  
  \smallskip
  
  In what follows, we will 
    investigate the statistical properties of the spatial distribution 
  of galaxies.
%
In the widely accepted model of galaxy formation, known as the {\it halo model}, galaxies form in the potential
wells of gravitationally collapsed  dark matter {\it halos}
 \cite{White:1977jf}.
  The 
distribution of galaxies naturally follow the distribution of halos.
   There are various approaches to characterize their formation, and how they are distributed in  space.  
In the  barrier crossing approach, initiated by  Press and Schechter \cite{Press:1973iz},
   halos are thought to form when the matter linear overdensity field overcomes  a certain
     threshold $\delta_c$. 
     We denote the
number of halos of mass $M$ at time $t$  by $n_h(\vec{x}, M,t)$:
%
 the Press-Schechter mass function gives the mean number density $n_h$ of halos as a function of the
  properties of the 
  underlying  linear matter  overdensity field $\delta_M$. In the following, we will drop the dependence on $M$ and $t$ (or $z$) to simplify the notation.
 
 In our work, 
    we  assume that
 $n_h(\vec{x})$  
 depends  on local physics only, i.e.
  it  is 
  a local function of $\delta(\vec{x})$  and of  its $n$-pt 
 correlation functions 
$$ 
\left[ \delta^n\right]_v (\vec{x})\,\equiv\,
 \int_{{k}_i\ge \ell^{-1}} \,\frac{d^3 k_1}{(2\pi)^3} \dots \frac{d^3 k_n}{(2\pi)^3}   
 \,e^{-i \left( \vec{k}_1+\dots+\vec{k}_n \right)\,\vec{x}}\,\langle \delta_{\vec{k}_1}\dots \delta_{\vec{k}_n}  \rangle
 $$ evaluated within a region $v\sim \ell^3$ around the position $\vec{x}$ (see Section \ref{sec-ss} for
an  additional
 discussion on this assumption):
 \be
 n_{h} (\vec{x})\,=\, \bar{n}_h (\delta(\vec{x}), \,\left[ \delta^n\right]_v (\vec{x}))\,.
 \ee  
 The 
   autocorrelations  $\left[ \delta^n\right]_v$ depend on the primordial
  intrinsic nG
    of the fundamental fields that seed the gravitational potential $\Phi$.  By measuring 
    how the statistics of the halo
    displacement 
    $$\delta_h=(n_h-\langle n_h \rangle)/\langle n_h \rangle$$
     is related to the statistics of the 
       matter density displacement 
        $\delta$, one can   extract information
 about primordial nG.
 
  The local statistics of the autocorrelations $\left[ \delta^n\right]_v$  can be calculated using a heuristic
 but powerful method, the 
 so called  peak-background split approach \cite{Bardeen:1985tr, Cole:1989vx}.   
 Consider a large subvolume $v\sim \ell^3$ of the universe containing a large number of halos. Take for concreteness 
  a fiducial scale  $\ell$, say 
  $\ell\,\sim\,10\,  {\rm Mpc}/h$, still much larger than the scale associated with halos of mass $M$.
  The linearized matter overdensity  field $\delta(\vec{x})$ 
can be decomposed into a short wavelength part $\delta_{short}$ and a long wavelength background $\delta_{long}$ 
with respect to the scale $\ell$: 
\be\label{sldec}
\delta(t,\,\vec{x})\,=\,\delta_{short}(t,\,\vec{x})+\delta_{long}(t,\,\vec{x})
\ee
where the short and long wavelength contributions to $\delta(t, \vec{x})$ are defined as
\bea
\delta_{short}(t,\,\vec{x})&=&\int_{k \ge \ell^{-1}}\frac{d^3 k}{(2 \pi)^3} \,e^{-i \vec{k} \vec{x}} \delta_{\vec{k}} (t)\,\\
\delta_{long}(t,\,\vec{x})&=&\int_{k \ll \ell^{-1}}\frac{d^3 k}{(2 \pi)^3} \,e^{-i \vec{k} \vec{x}} \delta_{\vec{k}} (t)\,.
\eea
 The short mode $\delta_{short}$ at scales $\le 10\, {\rm Mpc}/h$  can be considered as the source of  halos within the subvolume $v$ we are
  focussing
 on. As we mentioned above,  locally  $n_h$ depends on the quantity $\delta_{short}$ within the subvolume $v$, and on its correlation functions
 $\left[ \delta^n_{short}\right]_v$ evaluated on the same subvolume. 
  Instead, 
 distances larger than $100 {\rm Mpc}/h$, associated with long mode $\delta_{long}$, 
  are the scales over which we measure the statistics of halo displacements $\delta_h$. Within the subvolume
 $v$, 
 $\delta_{long}$ acts as   a smooth background density field  in the linear regime. 
 As we will discuss in the next sections,
 in the presence of primordial nG,  $\delta_{long}$ 
     has the important feature to modulate in a specific way the correlation functions  $\left[ \delta^n_{short}\right]_v$ 
    controlling   the  halo displacement $\delta_h$.
   %
  %
 This implies that  primordial nG
  of the gravitational potential  controls  the
 statistics of $\delta_h$ in such a way that  it makes   possible to extract from the latter unambiguous information on the former.

  Hence, the barrier crossing approach and the peak-background split  method
  allow us to analyze how
   primordial nG 
    affects the statistics of LSS.
 The connections
 between the two methods have been carefully discussed
  in  \cite{Baumann:2012bc,Ferraro:2012bd}: the statistics of 
 $\delta_{long}$ has the effect  to perturb the critical threshold that the linear part of $\delta_{short}$ has to reach to start to collapse, and
 it is this effect that renders the barrier crossing approach sensitive to primordial nG.

\smallskip

In the next sections, we will develop and apply these ideas to find quantitative results for the statistics of halo and matter
displacements, both at the level of power spectrum and bispectrum, for single and multiple sources. 

\section{The power spectrum
and the bispectrum: the single source case}\label{sec-ss}
 
 \noindent
 We start analyzing the case of single source, extending to the case of the bispectrum the results of \cite{Baumann:2012bc}.   
 We consider a large subvolume of the Universe containing many halos, over which the long mode is reasonably constant,
 and we denote with  $\left[ \cdot \right]_{\rm v}$ a spatial average over this subvolume -- while $\langle \cdot \rangle$ is the spatial
 average over the entire universe.
  In this section 
  we consider a single Gaussian field $\varphi$
  as an unique source for the primordial gravitational potential. 
 We  split such a field in long and short modes with
 respect to a fiducial scale $\ell$:
 \be\label{phisplit}
 \varphi\,=\,\varphi_s+\varphi_L\,.
 \ee
 This Gaussian field has
 zero average over the entire space $\langle \dots \rangle$, while non-zero average over the subvolume $\left[ \dots \right]_v$,
 with $v\sim \ell^3$:
 \bea
 \langle \varphi \rangle&=&0\,,\\
 \left[ \varphi \right]_{v}&=&\varphi_L\,. 
 \eea
 We 
 make a local expansion of  the primordial gravitational potential
  up to third order 
  in terms
 of the Gaussian field $\varphi$:
  \bea
 \Phi&=& \varphi
 +{\fnl} \left(\varphi^2-\langle \varphi^2\rangle\right)
+{\gnl}
 \left(
  \varphi^3
  -3 \langle \varphi^2\rangle \varphi
  \right)\,.
 \eea
 After implementing the long/short splitting we introduced in eq. (\ref{phisplit}), we get in coordinate space
 \bea
 \Phi&=& 
  \vl
 +{\fnl} \left(\vl^2-\langle \vl^2\rangle\right)
 +{\gnl}
 \left(
  \vl^3
  -3 \langle \vl^2\rangle \vl
  \right)
  \nonumber
\\&&
 +
 \left(1 +  2 \fnl \vl + 3 {\gnl} (\vl^2-\langle \vl^2\rangle)
 \right)\varphi_s
 \nonumber
 \\
&& +\left(\fnl+3 \gnl \vl
\right) \left(\varphi_s^2-[ \varphi_s^2]_v\right)\nonumber\\
 &&+
 \gnl
 \left(
  \varphi_s^3
  -3 [ \varphi_s^2 ] \varphi_s
  \right)\,.
 \eea
 The previous expression demonstrates
  how the long wavelength mode modulates the gravitational potential. The coefficients of the different powers of $\varphi_s$, 
  which control the statistics of this quantity within the subvolume $v$, receive contributions depending on the long mode $\varphi_L$,
   and this acts as a background quantity from the point of view of the subvolume $v$. In light of this, we can
 define the small scale effective power $ \sigma^e\,=\,\left[  \Phi^2\right]_v^{1/2}$, $\fnl^e$ and $\gnl^e$ as
 \bea\label{siges1}
\left( \sigma^e\right) &=&\left[ \vs^2 \right]_v^{1/2}\,
 \left(1 +  2 \fnl \vl +3{ \gnl} (\vl^2-\langle \vl^2\rangle) 
 \right)\,,
    \\ \label{fnles1}
    \fnl^{e}&=&\fnl+3 \gnl \vl \,,
    \\ \label{gnles1}
    \gnl^e&=&\gnl\,,
     \eea
 so we learn that long wavelength modes and primordial nG affect  the
 two point statistics of the short modes, as well as the non-Gaussian parameters inside the small box.
  The long mode dependence of eqs (\ref{siges1})-(\ref{gnles1}) 
  will be used to estimate the number of halos
  at a given position.
 
 \smallskip
 
 
  There are various approaches to the problem of quantifying how long modes contribute
  to  the process of halo formation.   The conceptually simplest one is to use an
univariate approach, in which the number of halos is expressed in power series of $\delta$, the matter density contrast. 
An univariate
  approach is 
 used  by Jeong and Komatsu \cite{Jeong:2009vd}.   
 This method includes the effects of the long mode   $\varphi_L$
 of the field sourcing the gravitational potential in the halo distribution, as much as $\varphi_L$ modulates $\delta$.   
 A physical issue with this approach is that, strictly speaking,
   the number of halos does not depend only  on the local value of the matter density contrast at a given  position $\vec{x}$. 
    Indeed, 
   it also depends on how the matter density contrast is {\it distributed} around $\vec{x}$ -- that is, on
its correlation functions evaluated at  $\vec{x}$. In the presence of local nG, these correlation functions are affected by the 
long mode  $\varphi_L$ of the primordial field that sources the gravitational potential (see eqs (\ref{siges1}) and (\ref{fnles1})), in a way that is not completely described by the dependence of $\delta$ on
  $\varphi_L$.
 A way to deal with this issue was
  adopted  by Giannantonio and Porciani \cite{Giannantonio:2009ak}:
  they propose  a bivariate approach, in which the halo number density is 
 expressed as power series in the long modes of both the matter density contrast {\it and} the gravitational potential.
 Following this method, 
 they 
show that  the results
   fit better with 
 N-body simulations. This approach has been then used
 by Baldauf et al in \cite{Baldauf:2010vn} for studying the features of galaxy bispectra.~\footnote{See also
 \cite{Yokoyama:2013mta} for a new approach, appeared as preprint while this work was being finalized.}   
 Here, we implement   another method that
 allows us to overcome
 the conceptual difficulty outlined above.   It is a generalization of the
 univariate approach that
   was introduced in
  \cite{Baumann:2012bc} in the context halo power spectra: we extend it to the study
  of bispectra.   We assume
 that the number of halos at a given position, $n_h(\vec x)$ does not depend on the matter density contrast only, but also on all its 
  correlation functions
   \be
  \label{ansnhc}
  n_h\,=\,n_h \left(\delta(\vec x),\,[ \delta^n]_v (\vec x) \right)\,.
  \ee 
 where $[ \delta^n]_v (\vec x) $ denotes the   correlation function $\langle \delta_1 \dots \delta_n\rangle$ evaluated over
 a volume of size $v$ around the point $\vec{x}$.
 Expressing the halo number as a function of the correlation functions of matter density contrast, we   
  faithfully 
 capture
 the contributions of the long mode $\varphi_L$ on the process of halo formation.
%
 %
      In a sense, the method that we adopt is a generalization of the
    univariate approach; on the other hand, it is also able to include features associated with the long 
     modes and primordial nG, like in the bivariate approach.  
     In order to analyze the properties induced by primordial nG only, 
     we  perform  a Taylor expansion of  the halo density contrast at first
   order in each of the arguments of $n_h$  in eq (\ref{ansnhc}). Further terms in the expansion
   would induce further non-linearities on the long mode dependence of $n_h$,
    and thus rely on the non-linear dependence of this function
    on the matter density contrast and its correlations. Since 
    we intend to focus our attention on the specific non-linearities associated with primordial nG only, 
     and we do not have full theoretical control over  higher derivatives of $n_h$ along the matter contrast correlation functions, 
    we discard these contributions,
    although it would be interesting to include them in a more complete analysis (see however section
    \ref{sec:gnlLSS} for some discussion on this point).
For the same reason,  
we do not include the effects of non-linear gravitational clustering in our analysis: that is, we work in Lagrangian space and linearly transform to Eulerian space. 
   As we will see, our procedure leads to 
manageable expressions  for the bispectra, that we use to derive interesting physical consequences; we will also compare
our results with others in the literature.

 \smallskip

 Let us now concretely apply the method we explained.
 We  are  assuming  that the halo number density $\left( n_h \right)_\ell$ within the subvolume is a function 
 of the one-point PDF of the underlying dark matter field $\delta_M$, linearly evolved until today and smoothed
 over the halo scale. This matter field is proportional, via the $W_M$ and $\alpha$ functions, to the gravitational
 potential $\Phi$ within the halo region: $\delta_M\,=\,\alpha\,\Phi$.   
 We define the quantity $\delta_L$ corresponding to 
 the
  primordial Gaussian long mode   as
 \be
 \delta_L\,=\, \alpha\,\varphi_L
\,, \ee
that is, in terms of the linearly evolved long mode for the Gaussian scalar field $\varphi_L$ (defined by integrating over small momenta
 the Fourier transform of $\varphi$).
  We emphasize that, consistently with the barrier crossing approach, $\delta_L$ and $\delta_M$ are proportional to the 
linearly evolved primordial fluctuations.
At very large scales, 
we can apply the transfer function to the Poisson equation to  express $\delta^{(L)}_{M}$ as \footnote{As we are working in the Fourier space, from now on powers of $\varphi_L$ should be intended as convolutions. We do this in order to keep a compact notation.}
  \be\label{expdM}
  \delta^{(L)}_{M}\,=\,\delta_L+\alpha\,\left[{\fnl}\,
 \left({\varphi_L^2}-{\langle\varphi_L^2\rangle}\right)
  + {\gnl}\,
\left( \varphi_L^3-3 \varphi_L\langle\varphi_L^2\rangle\right)\right] 
\,.
  \ee
Notice that $\delta^{(L)}_{M}$ is made by a combination of powers of long wavelength (Gaussian) modes. 
   For weakly non-Gaussian states, the one-point PDF in each subvolume can be characterized
 by its mean, 
 variance,
   and higher
 connected  cumulants. 
 Therefore, at very large
 scales, we can
 write an expansion
 for the halo
  displacement field  
  
 \bea\label{expfdh}
 \delta_h&=&b_g\,\left[
 \dl
 +{\alpha\,\fnl} \left(\vl^2-\langle \vl^2\rangle\right)
 +{\alpha\,\gnl}
  \left(\vl^3-3 \vl  \langle \vl^2\rangle\right)
 \right]\nonumber\\
 &&+\frac{\beta_2}{2} \left( \frac{\sigma_M^e}{
 \left[ \sigma_M^2 \right]_v^{1/2}
 } -
 1
 \right)+ \frac{\beta_3}{3}\,\left(\fnl^e-\fnl\right)+ \frac{\beta_4}{4}\,\left(\gnl^e-\gnl\right)\,
 \eea
  with the standard bias evaluated at ($\delta^{(L)}_{M}=0,\fnl,\gnl$)  by taking derivatives
  along the long wave-length mode $\delta^{(L)}_{M}$
  \be
  \label{eq:bg}
  b_g\,=\,\frac{\partial \ln n_h}{\partial \delta^{(L)}_{M}}\,.
  \ee
 Moreover, we define at the same point 
  \bea
  \label{eq:beta}
  \beta_2&=&2\, \frac{\partial \ln n_h}{\partial \ln \sigma_M^e} \hskip0.5cm,\hskip0.5cm\beta_3\,=\,3\,\frac{\partial \ln n_h}{\partial \fnl^e}
   \hskip0.5cm,\hskip0.5cm
  \beta_4\,=\, 4\, \frac{\partial \ln n_h}{\partial \gnl^e}
 \,. \eea
 A treatment similar to the previous one has been carried on
 also in \cite{Scoccimarro:2011pz}, where the derivatives are done directly in terms of correlation functions instead
 of nG parameters. 
  The quantities (\ref{eq:bg})-(\ref{eq:beta}) depend on the specific halo mass function one considers: we will discuss   
  theoretical expressions
  for the $\beta_i$ coefficients in section \ref{sec:gnlLSS}
  using 
  an extended Press-Schechter approach based on 
  an Edgeworth expansion. In the following analysis, we leave these parameters free.

  Using  formulae
  (\ref{siges1})-(\ref{gnles1}), eq. (\ref{expfdh}) becomes
  \bea
  \label{eqn:deltah}
  \delta_h&=& \delta_L\left(b_g+
   \frac{ \beta_2 \fnl}{\alpha} + \frac{ \beta_3 \gnl}{\alpha} 
   \right)
   \\
   &&+\left({\vl^2}-{\langle\vl^2\rangle}\right)\,\left(\,b_g\,\alpha \,\fnl+ \frac{3}{2} \beta_2 \,\gnl
   \right) 
   \\
   &&
   +\,\left( \vl^3-3 \vl \langle\vl^2\rangle\right) \,\left(  b_g \,\alpha\,  \gnl
   \right)  
 \,.
  \eea
 %
 %



 We can collect these results in the convenient expression
 \be\label{coldh}
  \delta_h\,=\,
  {b_1}{}\,\delta_L+\frac{b_2}{2}\,\left({\varphi_L^2}-{\langle\varphi_L^2\rangle}\right)
   + \frac{b_3}{6}\,\left( \varphi_L^3-3 \varphi_L\langle\varphi_L^2\rangle\right) 
 \,,
 \ee
 where we define the new bias parameters $b_i$ 
   as
   \bea
  b_1&\equiv&
  \,b_g+ \frac{\beta_2 \fnl}{\alpha }+ \frac{\beta_3 \gnl}{\alpha} 
   \label{eq:b1}\,,\\
   b_2&\equiv&
  2  \alpha\,b_g \,\fnl+
   3 \beta_2 \,\gnl
    \label{eq:b2}\,,
   \\
   b_3&\equiv& 
   6 \alpha\,b_g \gnl
   \label{eq:b3}\,.
   \eea
 The quantity $b_1$ corresponds to the linear bias. It receives a  contribution
 due to primordial nG, that scales as $1/\alpha\sim1/k^2$ at large scales: this is the well-known
 scale dependence of halo bias due to primordial nG. 
 In this large scale limit, the linear bias function $b_1$ depends on a particular combination of $\fnl$ and $\gnl$, and cannot 
 distinguish between these two quantities. On the other hand, the bias $b_2$ depends specifically on $\gnl$.
In what follows, we study the bispectrum of halos and matter, which depend on $b_2$: this can provide unambiguous  
 information on $\gnl$ allowing us to distinguish it from $\fnl$.

\smallskip

Before continuing, we
 re-emphasize that in our approach we define the linear and non-linear bias
 in eq. (\ref{coldh}) 
 as coefficients of powers of the primordial, linear Gaussian field $\varphi_L$: thus, $\delta_h$ acquires a non-Gaussian
 statistics of purely local form, and this allows straightforward computation of $n$-point functions. The primordial quantity $\vp_L$
  is then linearly evolved in terms of the function $\alpha(k)$, consistently with the barrier crossing approach in
  which   it is the linearly evolved gravitational potential that affects the formation of halos. Within this framework, as we will see
  in what follows, one can easily derive analytic expressions for the coefficients $\beta_i$ used above, using an Edgeworth expansion. 
  On the other hand, our approach differs from the univariate method of Jeong and Komatsu \cite{Jeong:2009vd} (see
  also \cite{Nishimichi:2009fs}), in which $\delta_h$ 
  is expressed in terms of powers of the evolved matter field with its own non-Gaussian features
  associated with non-linear evolution: 
     we will discuss more in details in
     Appendix \ref{app:comparison}  
     the differences among the results.   Nevertheless, our set-up has the virtue of being particularly intuitive  and simple to deal with, 
   and
   is fully consistent with the hypothesis at the basis of the barrier crossing approach that we adopt, that 
   it  is the linearly evolved matter density contrast  that determines  the formation of halos.
 As discussed above,   our generalized univariate approach  and results are more similar in spirit 
     to the 
     bivariate 
approach to bispectra adopted by Baldauf et al \cite{Baldauf:2010vn}: we will also compare in Appendix
\ref{app:comparison}
their results with our own.  

  \subsection{The two point function of halo and matter densities}
We adopt the following definitions for the
 power spectra associated with the 
 two point functions of halo and matter densities 
 \bea
  P_{mm}\,(2\pi)^3\,\delta(\vec{k}+\vec{p})&=&\langle \delta_M \delta_M\rangle\,
    \,,\\
  P_{hm}\,(2\pi)^3\,\delta(\vec{k}+\vec{p})&=&\frac{\langle \delta_h \delta_M\rangle + \langle \delta_M \delta_h\rangle}{2}
  \,,\\
  P_{hh}\,(2\pi)^3\,\delta(\vec{k}+\vec{p})&=&\langle \delta_h \delta_h\rangle\,
  \,.\eea
  We define the halo-halo power spectrum
  assuming that the shot-noise contribution $1/n_h$ has been subtracted, and analogously for the matter-halo
   and matter-matter power spectrum.  
  Using the formulae (\ref{expdM}), (\ref{coldh}) and the linear, primordial power spectrum\footnote{For simplicity we assume the scalar index $n_s=1$. } $\langle \varphi_L \varphi_L\rangle \,\equiv\, \Delta_0/k^3$, we find 
 \bea
  P_{mm}\,&=&
  \frac{\Delta_0\,\alpha^2}{k^3}
\,, \\
  P_{hm}\,&=&
  \frac{b_1\,\Delta_0\alpha^2}{k^3}
   \,,\\
  P_{hh}\,&=&
  \frac{b_1^2\,\Delta_0\,\alpha^2}{k^3}
  \,.\eea
  

 In the limit of very large scales, we can express the previous expressions for the power spectra as
  \bea
  P_{mm}\,&=&{\Delta_0\,\alpha_0^2 k}
 \,,\\
  P_{hm}\,&=&{k\,\left(\alpha_0\,b_g+\frac{\beta_2 \,\fnl}{k^2}+ \frac{\beta_3 \gnl}{k^2}\right)\,\Delta_0 \alpha_0 }
   \,,\\
  P_{hh}\,&=&{k\,\left(\alpha_0\,b_g+\frac{\beta_2 \,\fnl}{k^2}+ \frac{\beta_3 \gnl}{k^2}\right)^2\,\Delta_0 }
  \,,\eea
neglecting the terms that are subleading in powers of $k$, 
and using
 eq. (\ref{formulal})
which also holds specifically at large scales. Notice the famous scale dependence of halo bias induced by primordial nG
in the halo-halo power spectrum. 
  On the other hand, the  primordial non-Gaussian parameters $\fnl$ and $\gnl$ appear
   in the same footing in the previous expressions, rendering them 
   difficult to disentangle.
    In order
     to overcome this degeneracy, 
    in the following  we will consider the statistics of three point function and study the bispectra of halo and matter
    density contrasts. 
   
   
\subsection{The three point functions  of halo and matter densities}
  The three point functions are defined as \footnote{Note that $B_{hmm}$ stands for $(B_{hmm}+B_{mhm}+B_{mmh})/3$. The same for $B_{hhm}$. As for the power spectra, we assume that shot-noise contributions are removed.}
 
 \bea
  B_{mmm}\,(2\pi)^3\,\delta(\vec{k}_1+\vec{k}_2+\vec{k}_3)&=&\langle \delta_M \delta_M \delta_M\rangle\,
 \,,\\
  B_{hmm}\,(2\pi)^3\,\delta(\vec{k}_1+\vec{k}_2+\vec{k}_3)&=&\frac{\langle \delta_h \delta_M \delta_M\rangle + \langle \delta_M \delta_h \delta_M\rangle + \langle \delta_M \delta_M \delta_h \rangle}{3}\,
   \,,\\
  B_{hhm}\,(2\pi)^3\,\delta(\vec{k}_1+\vec{k}_2+\vec{k}_3)&=&\frac{\langle \delta_h \delta_h \delta_M\rangle + \langle \delta_h \delta_M \delta_h \rangle + \langle \delta_M \delta_h \delta_h \rangle}{3}\,
    \,,\\
  B_{hhh}\,(2\pi)^3\,\delta(\vec{k}_1+\vec{k}_2+\vec{k}_3)&=&\langle \delta_h \delta_h \delta_h\rangle\,
 \,, \eea

  obtaining at leading order (tree-level)

   \bea
  B_{mmm}&=&\,2 \fnl\Delta_0^2\, \alpha^{(1)}\alpha^{(2)} \alpha^{(3)}\left(\frac{1}{k_1^3 k_2^3}+\frac{1}{k_1^3 k_3^3}+\frac{1}{k_2^3 k_3^3}\right)\label{bsmmm}
 \,,\\
  B_{hmm}&=&\,\frac{1}{3}\left\{\Delta_0^2\alpha^{(2)}\alpha^{(3)}\left[2 \fnl\frac{\alpha^{(1)}\,b_1^{(1)}}{k_1^3}\right(\frac{1}{k_2^3}+\frac{1}{k_3^3}\left)+\frac{b_2^{(1)}}{k_2^3 k_3^3} \right]+\text{perm.}\right\}\label{bshmm}
  \,, \\
  B_{hhm}&=&\,\frac{1}{3}\left\{\Delta_0^2\alpha^{(3)}\left[2\fnl \frac{\alpha^{(1)}\,b_1^{(1)}}{k_1^3}\frac{\alpha^{(2)}\,b_1^{(2)}}{k_2^3}+\frac{1}{k_3^3}\left(b_2^{(2)}\frac{\alpha^{(1)}\,b_1^{(1)}}{k_1^3}+b_2^{(1)}\frac{\alpha^{(1)}\,b_1^{(2)}}{k_2^3}\right)\right]+\text{perm.}\right\}\label{bshhm}
    \,,\\
  B_{hhh}&=&\,\Delta_0^2\left(b_2^{(1)}\frac{\alpha^{(2)}\,b_1^{(2)}}{k_2^3}\frac{\alpha^{(3)}\,b_1^{(3)}}{k_3^3}+b_2^{(2)}\frac{\alpha^{(1)}\,b_1^{(1)}}{k_1^3}\frac{\alpha^{(3)}\,b_1^{(3)}}{k_3^3}+b_2^{(3)}\frac{\alpha^{(1)}\,b_1^{(1)}}{k_1^3}\frac{\alpha^{(2)}b_1^{(2)}}{k_2^3}\right)\label{bshhh}
\,,  \eea
  where $\alpha^{(j)}\equiv\alpha(k_j,z)$ and $b_i^{(j)}\equiv\,b(k_j,z)$. Notice that the scale dependence of 
 $ b_i^{(j)}$ is due to  primordial nG, and this induces a dependence on $\alpha^{(j)}$ 
 for these quantities 
 (see eqs. (\ref{eq:b1})-(\ref{eq:b3})).
    In the previous formulae, we focussed only on tree-level 
  contributions to the bispectra,
    neglecting so-called ``loop'' effects. For this reason,  the bias parameter $b_3$ does not appear.  Focussing 
    on the halo bispectrum, eq. (\ref{bshhh}), recall that the bias parameters $b_1$ and $b_2$ depend 
    on primordial nG,  see eqs (\ref{eq:b1},\,\ref{eq:b2}). We are thus able to recover the trispectrum dependent
    contributions identified by Jeong and Komatsu \cite{Jeong:2009vd}. We will explain in 
    Appendix \ref{app:comparison} how to compare
      (in appropriate limits) our results with \cite{Jeong:2009vd}: for the moment, we will elaborate the
     physical consequences of our findings.

\subsubsection{Isosceles triangles and the squeezed limit in momentum space}
\label{subsec:squeezed}

  We focus on isosceles triangle configurations, considering
  a squeezed limit in which  
 primordial local nG contributions yield the largest signal \cite{Jeong:2009vd}.
  We now follow the convention of Jeong and Komatsu by setting $k=k_1=k_2=\epsilon k_3$: in other words, 
   we consider  
    isosceles triangles in momentum space. The squeezed limit, then, corresponds to configurations
 in which $\epsilon\gg1$.
 The previous expressions (\ref{bsmmm})-(\ref{bshhh}) become
%

   \bea
  B_{mmm}&=&\,\frac{2 \fnl\Delta_0^2\, \alpha(k) \alpha(k/\epsilon) \,\left(1+\epsilon^3\right)}{k^6} 
\label{bsqmmmMS}
 \,,\\
  B_{hmm}&=&\,\frac{\Delta_0^2\,\alpha(k)}{3\,k^6}\left\{
  4 \fnl \left[\left( 1+ \epsilon^3\right)\,\alpha\left(\frac{k}{\epsilon}\right) \,\alpha(k)\,b_1(k)
  +\epsilon^3 \,\alpha(k) \,\alpha(\frac{k}{\epsilon})\,b_1\left(\frac{k}{\epsilon}\right)
  \right]+
  \left[ 2 \epsilon^3\,\alpha\left(\frac{k}{\epsilon}\right) \,b_2(k)+  \alpha(k) \,b_2\left(\frac{k}{\epsilon}\right)\right]
  \right\}\nonumber\\
  \label{bsqhmmMS}
   \,,\\
  B_{hhm}&=&
\frac{\Delta_0^2}{3\,k^6}\Big\{
2 \fnl \left[\,\alpha\left(\frac{k}{\epsilon}\right) \,\alpha(k)^2\,b_1(k)^2
  +2\,\epsilon^3 \,\alpha(k)^2\,\alpha\left(\frac{k}{\epsilon}\right) \,b_1(k)\,b_1\left(\frac{k}{\epsilon}\right)
  \right]+\nonumber\\
  &&+
  \left[ 2 \epsilon^3\,\alpha\left(\frac{k}{\epsilon}\right)\,\alpha(k)\,b_1(k) \,b_2(k)+ 2 \alpha(k) \left(\epsilon^3
  \,\alpha\left(\frac{k}{\epsilon}\right)\,b_1\left(\frac{k}{\epsilon}\right) b_2(k)+ b_2\left(\frac{k}{\epsilon}\right) \alpha(k)\,b_1(k)\right)\right]
\Big\}
  \label{bsqhhmMS}
    \,,\\
  B_{hhh}&=&\,
  \frac{\Delta_0^2\,\alpha(k)\,b_1(k)}{k^6}
\left\{ 2 \epsilon^3
\,\alpha\left(\frac{k}{\epsilon}\right) \, b_1\left(\frac{k}{\epsilon}\right) b_2(k)+ b_2\left(\frac{k}{\epsilon}\right) \alpha(k)\,b_1(k)
  \right\}
  \,,\label{bsqhhhMS}
  \eea
where we keep all contributions. 
 These results, which use our generalized univariate approach, are similar to 
 the ones obtained by  \cite{Baldauf:2010vn} using 
a bivariate approach: see Appendix \ref{app:comparison} for more details.  Expanding the previous functions,
 one finds at large scales $k\ll1$ and in squeezed limit
 $\epsilon\gg1$, using $\alpha(k)\,=\,\alpha_0\,k^2$ and the approximation $T(k/\epsilon)\simeq T(k)$,

   \bea
  B_{mmm}&\simeq&\,4 \fnl \Delta_0^2 \alpha_0^3 \epsilon\,,
 \\
  B_{hmm}&\simeq&\,4 \fnl \Delta_0^2 \alpha_0^3 b_g \epsilon + \frac{2 \Delta_0^2 \alpha_0^2}{3k^2}\left[(\beta_2 \left(3 \gnl \epsilon + 2 \fnl^2 \epsilon^3\right) + 2 \fnl \gnl \beta_3 \epsilon^3 \right]\,,
     \\
  B_{hhm}&\simeq&\,4 \fnl \Delta_0^2 \alpha_0^3 b_g^2 \epsilon + \frac{4 \Delta_0^2 \alpha_0^2 b_g}{3k^2} \left[\beta_2 \left(3 \gnl \epsilon + 2 \fnl^2 \epsilon^3\right) + 2 \fnl \gnl \beta_3 \epsilon^3\right]\,,
 \nonumber\\
 && +\frac{2 \Delta_0^2 \alpha_0 \epsilon^3}{3k^4}\left[\beta_2^2 \left(2 \fnl^3 + 3 \fnl \gnl \right) + \beta_2 \beta_3 \left(4 \fnl^2 \gnl + 3 \gnl^2 \right)+2 \fnl \gnl^2 \beta_3^2 \right]\,,
    \\
  \label{eq:bhhhk}
  B_{hhh}&\simeq&\,4 \fnl \Delta_0^2 \alpha_0^3 b_g^3 \epsilon + \frac{2 \Delta_0^2 \alpha_0^2 b_g^2}{k^2}\left[\beta_2 \left(3 \gnl \epsilon + 2 \fnl^2 \epsilon^3\right) + 2 \fnl \gnl \beta_3 \epsilon^3 \right]\,,
  \nonumber\\
  && +\frac{2\alpha_0\Delta_0^2 b_g \epsilon^3}{k^4}\left[\beta_2^2 \left(2 \fnl^3 + 3 \fnl \gnl \right) + \beta_2 \beta_3 \left(4 \fnl^2 \gnl  + 3 \gnl^2 \right) + 2 \fnl \gnl^2 \beta_3^2 \right]\,,
  \nonumber\\
  && +\frac{6 \gnl \Delta_0^2 \beta_2 \epsilon^3}{k^6} \left(\fnl \beta_2 + \gnl \beta_3 \right)^2\,.
  \eea 
The expansion of the various expressions for the bispectra at large scales makes 
manifest  how the non-Gaussian parameters $\fnl$ and $\gnl$ characterize different bispectra, and suggests
  that the study of bispectra allows to distinguish the effects of each of them. Indeed, several contributions appear with different
  scale dependences and different coefficients: these can be used to distinguish the effects of $\fnl$ and $\gnl$.
  
  \subsection{Methods to disentangle $\gnl$ from $\fnl$}
  
  Here we discuss how our results allow one to break the degeneracy between  $\gnl$
  and $\fnl$ that affects the power spectra. 
    We will discuss two possible  methods to use bispectra to distinguish these two quantities. 

\smallskip

  The previous results are obtained in the limit of large scales. We use this limit in order to allow simple analytical approximations
  for the transfer function $T(k)$ and thus allow us to analytically understand physically interesting
  features of the bispectra as a function of the scale $k$, for configurations of squeezed isosceles triangles in momentum space
  $k=k_1=k_2=\epsilon k_3$.

    As a specific example,
      let us study in more detail the properties of $B_{hhh}$, using the eq. (\ref{eq:bhhhk}) valid at large scales.
       The aim is to determine distinctive signatures of $\gnl$ in the profile of halo bispectra.  
   While  
  eq. (\ref{eq:bhhhk}) has been written in a form aimed to emphasize the different scale dependences of the various
  contributions, it can also be expressed in a more concise form as 
  \be\label{bhhhz}
  B_{hhh}\,=\,\frac{2 \Delta_0 \epsilon}{k^6}\left(3\beta_2 \gnl+2\alpha_0 b_g\,\fnl\,k^2 \right)\,
  \left[\epsilon^2\left( \beta_2 \fnl+\beta_3 \gnl\right)^2+\alpha_0 b_g \epsilon^2\left( \beta_2 \fnl+\beta_3 \gnl\right)
  k^2+\alpha_0^2 b_g^2 k^4
  \right]\,,
  %
  %
  \ee
which makes the zeroes of $B_{hhh}$ clearly identifiable. Recall that the bispectrum (\ref{bhhhz}) is defined for isosceles configurations
in momentum space, in which the equal sides of the triangle have length $k$, while the small side length $k/\epsilon$. 
Hence, we are studying $B_{hhh}$ as a function of the size of the long sides of an isosceles triangle in momentum space. 
  In the limit of large $\epsilon$, eq.  (\ref{bhhhz})  has various roots: the ones that can be real are
  
  \bea
  k_{root}^{(1)}&=&\sqrt{-\frac{3\,\beta_2\,\gnl}{2\,\alpha_0\,b_g\,\fnl}}\label{kroot1}\,,\\
   k_{root}^{(2)}&=&\sqrt{\frac{\epsilon^2 \left| \beta_2 \fnl+\beta_3 \gnl  \right|}{\alpha_0\,|b_g|}}\label{kroot2}
 \,, \eea
  hence the existence and positions of the zeroes of galaxy bispectrum depend on the values of the non-Gaussian
  parameters $\fnl$ and $\gnl$. Working in regimes in which the parameters
  $\alpha_0$, $\beta_2$ and $b_g$ are positive (see the discussion
   in Section \ref{sec:gnlLSS}),   the quantity  $k_{root}^{(1)}$ is real only if $\fnl $
  and $\gnl$ 
   are both non-vanishing and have  opposite signs.  
      Hence, if the profile 
       of $B_{hhh}$  changes sign as a function of the scale, it would be a tantalizing hint 
      of the presence of non-vanishing $\gnl$ with an opposite sign to $\fnl$.  
   While these roots have been derived in the limit of large scales where we can neglect the scale dependence
   of the transfer function, we will show that these features are accurately reproduced by a full numerical analysis
   in Section \ref{sec:gnlLSS}.

  \smallskip
  
  The bispectrum $B_{hhh}$ is not the only quantity that allows to distinguish among different nG parameters.   
Combining different bispectra, indeed, one can more directly probe the individual effects of $\fnl$
with negligible contamination 
 from $\gnl$.
For example, let us consider the combination
%

  
   \be\label{intcomb}
    C_{fnl}\,\equiv\,\frac{B_{hhh}}{P_{hh}^2}+3 \frac{B_{hmm}}{P_{hm}^2}-3 \frac{B_{hhm}}{P_{hm} P_{hh}}\,.
    \ee
 {\it   Without} making any approximation, using equations (\ref{bsqmmmMS})-(\ref{bsqhhhMS})
  one finds that the previous quantity reads, for isosceles triangles, 
  \be\label{dcfnl}
     C_{fnl}\
    \,=\,\frac{2\fnl\,\left(1+2 \epsilon^3\right)\,\alpha\left( k/\epsilon\right)}{
    \alpha(k)\,\left(
 b_g\, \alpha(k) +
  \fnl \beta_2 + \gnl \beta_3 \right)
    }\,.
   \ee  
    Hence it depends specifically on $\fnl$ (with only a minor dependence on $\gnl$ in the denominator). Hence, a measurement
    of non-vanishing $C_{fnl}$ would provide a very clean probe of the quantity $\fnl$ that is (almost) independent  on the size of $\gnl$.
    Nevertheless, the combination
   $C_{fnl}$
    is  challenging 
   to probe observationally, since it is harder to observe bispectra and power spectra involving
   dark matter densities (although optimistically this might be realized in the future
   using gravitational lensing).


The bottom-line of this section is that the bispectra  of halos and matter have distinctive qualitative features that allow one to distinguish
the effects of different non-Gaussian parameters, thereby providing observables that may allow one to break the degeneracies between $f_{\rm NL}$ and $g_{\rm NL}$ that are present in the study of power spectra
only.  Let us emphasize once more that the analytical results obtained so far include  the effects of primordial nG only, without taking into account non-linear effects due to gravity. This approximation allowed 
 us   to analytically identify more directly the effects of different
nG parameters in the halo distribution. A systematic analysis 
 is needed to understand how much the features we pointed out survive when
including the effects of gravity, and is left for future work.  See however Section \ref{sec:gnlLSS} for a 
   numerical study of our analytic  results, with some preliminary discussion on possible effects of non-linear gravitational
   clustering.  
 The next two sections are more theoretical, and investigate the implications of our findings when applied to the multiple source extension of our treatment.

\section{The power spectrum
and the bispectrum: the multiple source case}
 \label{sec:npoint}
 
 \noindent
 In this section, we discuss how to extend the previous analysis to the case of more than one
 source field for the primordial gravitational potential. We focus for definiteness on
 the case of two fields, that contribute to the gravitational potential as in the following Ansatz
  \bea\label{mansatz}
 \Phi&=& \varphi+\psi
 +{{\fnl\,\left(1+\Pi\right)^2}} \,\left(\psi^2-\langle \psi^2\rangle\right)
+{\gnl \,\left(1+\Pi\right)^3}\,
 \left(
  \psi^3
  -3 \langle \psi^2\rangle \psi
  \right)\,,
 \eea
 with
\be
\Pi\,=\,\frac{P_{\varphi}}{P_{\psi}}\,.
\ee
While $\varphi$ can be thought as the inflaton fluctuation, $\psi$ can be seen as  a spectator field that is responsible
 for introducing nG of local form in the gravitational potential.
The two fields  $\varphi$ and $\psi$ are by themselves Gaussian: 
 we  split them  as long and short mode with
 respect to a fiducial scale $\ell$, as done in the previous section:
 \bea
 \varphi&=&\varphi_s+\varphi_L\,,
 \\
  \psi&=&\psi_s+\psi_L\,.
 \eea

This set-up can be seen as an extension of 
  the curvaton-like model
 analyzed in section of 4.1 of
 \cite{Baumann:2012bc}, where we include   the contribution of $\gnl$. From now on,
 $ P_i = P_\Phi (k_i)$ and $P_{ij}  = P_\Phi(| k_i + k_j |)$.
 Starting from Ansatz (\ref{mansatz}), the three and four pt functions read
 \bea
 \langle \Phi(k_1) \Phi(k_2) \Phi(k_3) \rangle&=& {\fnl} \left[ P_1 P_2+\text{5 perms }\right]\,,\\
  \langle \Phi(k_1) \Phi(k_2) \Phi(k_3) \Phi(k_4) \rangle&=&2\,\left(\frac56 \right)^2\,\taunl \left[ P_1 P_2 P_{13}+\text{23 perms }\right]
+{\gnl}\,  \left[ P_1 P_2 P_{3}+\text{11 perms }\right]\,,
 \eea
 where
 \be
 \taunl\,=\,\left(\frac65 \fnl \right)^2\,\left(1+\Pi\right)\,.
 \ee
The usual equality   $ \taunl\,=\,\left(\frac65 \fnl \right)^2$ is recovered, as expected, in the single field limit $\Pi \to 0$.

 After implementing the long/short splitting, one gets in coordinate space
 \bea
 \Phi&=& 
 \varphi_L+\psi_L
 +{{\fnl\,\left(1+\Pi\right)^2}} \,\left(\psi_L^2-\langle \psi_L^2\rangle\right)
+{6 \gnl \,\left(1+\Pi\right)^3}\,
 \left(
  \psi_L^3
  -3 \langle \psi_L^2\rangle \psi_L
  \right)
  \,
  \nonumber
\\&&
 +\varphi_s+
 \left(1 +  2 \fnl \,\left(1+\Pi\right)^2 \, \psi_L + 3 {\gnl} \,\left(1+\Pi\right)^3\,(\psi_L^2-\langle \psi_L^2\rangle)
 \right)\psi_s\,
 \nonumber
 \\
&& +{\left(1+\Pi\right)^2}\left( \fnl+3\,\left(1+\Pi\right)\,\gnl \psi_L
\right) \left(\psi_s^2-[ \psi_s^2]_v\right)\nonumber\\
 &&+ 
 \gnl\,\left(1+\Pi\right)^3\,
 \left(
  \psi_s^3
  -3 [ \psi_s^2 ] \psi_s
  \right)
  \,.
 \eea
 The previous expression demonstrates
  how long wavelength modes modulate the gravitational potential. The coefficients of the different powers of $\varphi_s$
  and $\psi_s$,  
 which control the statistics of this quantity within the subvolume $v$, receive contributions depending on the long mode $\psi_L$,
   and thus acts as a background quantity from the point of view of the subvolume $v$. In  light of this, we can
 define the small scales effective power $ \sigma^e\,=\,\left[  \Phi^2\right]_v^{1/2}$, $\fnl^e$ and $\gnl^e$ as
 \bea \label{siges2}
\left( \sigma^e\right) &=&\left[ \Phi_s^2 \right]_v^{1/2}
\,\Big\{
1+2\,
  \fnl (1+\Pi) \psi_L +\left(1+\Pi\right)^2\left(2\,{\fnl^2}\,\Pi\,\psi_L^2+3\,{ \gnl} (\psi_L^2-\langle \psi_L^2\rangle) \right)
  \nonumber\\
&&+ \frac12\,\left(1+\Pi\right)^3\,
\Pi\,
\left( -8\,{\fnl^3}\,\psi_L^3+12\,{\fnl \gnl} \,\psi_L (\psi_L^2-\langle \psi_L^2\rangle) \right)
 \Big\}
    \,,\\ \left(1+\Pi\right)^2
    \fnl^{e}&=&\left(1+\Pi\right)^2\,
    \left(\fnl+3\,\left(1+\Pi\right)\,\gnl \psi_L
\right)
    \,,\\ \label{gnles2}
\left(1+\Pi\right)^3    \gnl^e&=&\left(1+\Pi\right)^3\,\gnl
    \,, \eea
 so we learn that long wavelength modes affect the
 two-point statistics of the short modes, as well as the non-Gaussian parameters inside the small box.
 
 
 \smallskip

Following our treatment of the single source case, we
 can define
  the quantity $\delta_L$ corresponding to the linearly evolved sum of primordial Gaussian perturbation  as
 \be
 \delta_L\,=\, \alpha\,\left(\varphi_L+\psi_L\right)
 \,.
 \ee

 
 Using the Poisson equation, the long modes in the smoothed matter density contrast reads

  \be
  \delta^{(L)}_{M}\,=\,\delta_L+\alpha\,\left[{\left(1+\Pi\right)^2\,\fnl}\,
 \left({\psi_L^2}-{\langle\psi_L^2\rangle}\right)
  +{\left(1+\Pi\right)^3\,\gnl}\,
\left( \psi_L^3-3 \psi_L\langle\psi_L^2\rangle\right)\right] 
  \,.\ee

 



We express the halo density contrast as the expansion

  
 \bea
 \delta_h&=&{b_g}\,\left[
 \dl
 +{\alpha\, \left(1+\Pi\right)^2\,\fnl} \left(\psi_L^2-\langle \psi_L^2\rangle\right)
 +{\alpha\, \left(1+\Pi\right)^3\,\gnl} 
  \left(\psi_L^3-3 \psi_L  \langle \psi_L^2\rangle\right)
 \right]\nonumber\\
 &&+\frac{\beta_2}{2}
  \left(
  \frac{{\sigma_M^e}}{{
 \left[ \sigma_M^2 \right]_v^{1/2}
 }}-1 
 \right)+ \frac{\beta_3}{3}\, \left(\fnl^e-\fnl\right)+\frac{\beta_4}{4}\, \left(\gnl^e-\gnl\right)
\,, \eea
  with the standard bias evaluated at $(\delta^{(L)}_{M}=0,\fnl,\gnl)$
  \be
  b_g\,=\,\frac{\partial \ln n_h}{\partial \delta^{(L)}_{M}}\,,
  \ee
  and, moreover, we define at the same point
  \bea
  \beta_2&=&2 \frac{\partial \ln n_h}{\partial \ln \sigma_M^e} \hskip0.5cm,\hskip0.5cm\beta_3\,=\,3 \frac{\partial \ln n_h}{\partial  \fnl^e} \hskip0.5cm,\hskip0.5cm\beta_4\,=\,4 \frac{\partial \ln n_h}{\partial\,\gnl^e}\,.
  \eea
  %
  Substituting the formulae above, one finds
   
  \bea
  \label{eqn:deltah2}
  \delta_h&=& b_g\, \delta_L +
   2\,\beta_2\, \left(1+\Pi\right)\, \fnl\,\psi_L + 6 \beta_3\, \gnl\, \left(1+\Pi\right) \,\psi_L  
   \nonumber\\
   &&+\frac{ \left(1+\Pi\right)^2}{2 }\left[\left({\psi_L^2}-{\langle\psi_L^2\rangle}\right)\,\left(2\,b_g\,\alpha \,\fnl+ 3\beta_2\,\gnl\right) + \psi_L^2\beta_2\,\fnl^2 \Pi\right]
    \nonumber\\
   &&
   +\,\frac{ \left(1+\Pi\right)^3}{6  }\left[\left( \psi_L^3-3 \psi_L \langle\psi_L^2\rangle\right) \,\left(6 b_g \,\alpha\, \gnl +18{\Pi}\,\beta_2\,\fnl\,\gnl\right)-24\Pi \fnl^3 \beta_2 \psi_L^3\right]
   \\
     &=&b_g\,\alpha\, \vp_L+b_1\,\alpha \,\psi_L+\frac{1}{2}\,\left({b_2 \psi_L^2}-b_3 {\langle\psi_L^2\rangle}\right)
   + \frac{1}{6}\,\left( b_4\,\psi_L^3-3 b_5\, \psi_L\langle\psi_L^2\rangle\right)\,.
  \eea
   We define the bias $b_i$    as 
   \bea\label{mb1}
  {b_1}&\equiv&
  b_g
   +
    \beta_2\, \left(1+\Pi\right)\, \frac{\fnl}{ \alpha} +  \beta_3\,  \left(1+\Pi\right)\,\frac{ \gnl}{\alpha}
  \,, \\ \label{mb2}
   b_2&\equiv&
   \left(1+\Pi\right)^2\,
  \left[2  \alpha\,b_g \,\fnl+
   3 \beta_2 \,\gnl + \Pi \beta_2 \fnl^2 
  \right]\,, \\
   b_3&\equiv& 
   \left(1+\Pi\right)^2\,
  \left[2  \alpha\,b_g \,\fnl +
   3 \beta_2 \,\gnl \right]\,, \\
   b_4&\equiv& 
   \left(1+\Pi\right)^3
   \left[
  3 b_g \,\alpha\,  \gnl
   +9 {\Pi}\,\beta_2\,\fnl\,\gnl 
   -12{\Pi}\beta_2 \fnl^3\right]
   \,, \\
   b_5&\equiv& 
   \left(1+\Pi\right)^3
   \left[
  3 b_g \,\alpha\,  \gnl
   +9{\Pi} \,\beta_2\,\fnl\,\gnl 
   \right]
 \,.  \eea

   \subsection{The two point functions of halo and matter densities}
We adopt the following definitions for the two point functions (the same as in the single source case 
we discussed in
previous sections)
 \bea
  P_{mm}\,(2\pi)^3\,\delta(\vec{k}+\vec{p})&=&\langle \delta_M \delta_M\rangle\,
   \,, \\
  P_{hm}\,(2\pi)^3\,\delta(\vec{k}+\vec{p})&=&\frac{\langle \delta_h \delta_M\rangle + \langle \delta_M \delta_h\rangle}{2}
  \,,\\
  P_{hh}\,(2\pi)^3\,\delta(\vec{k}+\vec{p})&=&\langle \delta_h \delta_h\rangle\,
  \,,\eea
where
 \bea
  P_{mm}\,&=&\,\frac{\Delta_{mm}}{k^3}\,=\,\frac{\Delta_0\,\alpha^2\,\left(1+\Pi\right)}{k^3}
 \,,\\
  P_{hm}\,&=&\,\frac{\Delta_{hm}}{k^3}\,=\,\frac{\left(\alpha\,b_1 +\alpha \,b_g\,\Pi\right)\,\Delta_0\alpha}{k^3}
   \,,\\
  P_{hh}\,&=&\,\frac{\Delta_{hh}}{k^3}\,=\,\frac{\left(\alpha^2\,b_1^2 +\alpha^2 \,b^2_g\,\Pi\right)\,\Delta_0}{k^3}
 \,. \eea
 
\subsection{The three point functions of halo and matter densities}
  The three point functions are defined as 
 
 \bea
  B_{mmm}\,(2\pi)^3\,\delta(\vec{k}_1+\vec{k}_2+\vec{k}_3)&=&\langle \delta_M \delta_M \delta_M\rangle\,
 \,,\\
  B_{hmm}\,(2\pi)^3\,\delta(\vec{k}_1+\vec{k}_2+\vec{k}_3)&=&\frac{\langle \delta_h \delta_M \delta_M\rangle + \langle \delta_M \delta_h \delta_M\rangle + \langle \delta_M \delta_M \delta_h \rangle}{3}\,
   \,,\\
  B_{hhm}\,(2\pi)^3\,\delta(\vec{k}_1+\vec{k}_2+\vec{k}_3)&=&\frac{\langle \delta_h \delta_h \delta_M\rangle + \langle \delta_h \delta_M \delta_h \rangle + \langle \delta_M \delta_h \delta_h \rangle}{3}\,
    \,,\\
  B_{hhh}\,(2\pi)^3\,\delta(\vec{k}_1+\vec{k}_2+\vec{k}_3)&=&\langle \delta_h \delta_h \delta_h\rangle\,
\,,  \eea

  obtaining

   \bea
  B_{mmm}&=&\,2\,\left( 1+\Pi\right)^2\,\fnl\Delta_0^2\, \alpha^{(1)}\alpha^{(2)} \alpha^{(3)}\left(\frac{1}{k_1^3 k_2^3}+\frac{1}{k_1^3 k_3^3}+\frac{1}{k_2^3 k_3^3}\right)\label{bmmmm}
 \,,\\
  B_{hmm}&=&\,\frac{1}{3}\left\{\Delta_0^2\alpha^{(2)}\alpha^{(3)}\left[4\,\fnl\,\left( 1+\Pi\right)^2\,\frac{\alpha^{(1)}\,b_1^{(1)}}{k_1^3}\right(\frac{1}{k_2^3}+\frac{1}{k_3^3}\left)+\frac{b_2^{(1)}}{k_2^3 k_3^3} \right]+\text{perm.}\right\}\label{bmhmm}
  \,,\\
  B_{hhm}&=&\,\frac{1}{3}\left\{\Delta_0^2\alpha^{(3)}\left[2\,\fnl\,\left( 1+\Pi\right)^2\, \frac{\alpha^{(1)}\,b_1^{(1)}}{k_1^3}\frac{\alpha^{(2)}\,b_1^{(2)}}{k_2^3}+\frac{1}{k_3^3}\left(b_2^{(2)}\frac{\alpha^{(1)}\,b_1^{(1)}}{k_1^3}+b_2^{(1)}\frac{\alpha^{(2)}\,b_1^{(2)}}{k_2^3}\right)\right]+\text{perm.}\right\}\label{bmhhm}
    \,,\\
  B_{hhh}&=&\,\Delta_0^2\left(b_2^{(1)}\frac{\alpha^{(2)}\,b_1^{(2)}}{k_2^3}\frac{\alpha^{(3)}\,b_1^{(3)}}{k_3^3}+b_2^{(2)}\frac{
  \alpha^{(1)}\,
  b_1^{(1)}}{k_1^3}\frac{\alpha^{(3)}\,b_1^{(3)}}{k_3^3}+b_2^{(3)}\frac{\alpha^{(1)}\,b_1^{(1)}}{k_1^3}\frac{\alpha^{(2)}\,b_1^{(2)}}{k_2^3}\right)\label{bmhhh}
  \,,\eea
  where $\alpha^{(j)}\equiv\alpha(k_j,z)$ and $b_i^{(j)}\equiv\,b(k_j,z)$.  In the previous formulae, we focussed only on tree-level 
  contributions, neglecting loop effects.   
  Remarkably, only the bias parameter $b_1$ and $b_2$ appear.

\subsubsection{Isosceles triangles and the squeezed limit in momentum space}\label{sec:isomulti}

We now set 
 $k=k_1=k_2=\epsilon k_3$. That is, we consider isosceles triangles. The squeezed limit, then, corresponds to configurations
 in which $\epsilon\gg1$.
 The previous expressions (\ref{bmmmm})-(\ref{bmhhh}) become
%

   \bea
  B_{mmm}&=&\,\frac{2\,\fnl\,\left( 1+\Pi\right)^2\,\Delta_0^2\, \alpha(k) \alpha(k/\epsilon) \,\left(1+\epsilon^3\right)}{k^6} 
\label{bsqmmm}
 \,,\\
  B_{hmm}&=&\,\frac{\Delta_0^2\,\alpha(k)}{3\,k^6}\Big\{
  4 \fnl \,\left( 1+\Pi\right)^2\,\left[\left( 1+ \epsilon^3\right)\,\alpha\left(\frac{k}{\epsilon}\right) \,\alpha(k)\,b_1(k)
  +\epsilon^3 \,\alpha(k) \,\alpha\left(\frac{k}{\epsilon}\right)\,b_1\left(\frac{k}{\epsilon}\right)
  \right]+\nonumber\\
  &&+
  \left[ 2 \epsilon^3\,\alpha\left(\frac{k}{\epsilon}\right) \,b_2(k)+  \alpha(k) \,b_2\left(\frac{k}{\epsilon}\right)\right]
  \Big\}
  \label{bmsqhmm}
   \,,\\
  B_{hhm}&=&
\frac{\Delta_0^2}{3\,k^6}\Big\{
2\,\fnl \,\left( 1+\Pi\right)^2\, \left[\,\alpha\left(\frac{k}{\epsilon}\right) \,\alpha(k)^2\,b_1(k)^2
  +2\,\epsilon^3 \,\alpha(k)^2 \,\alpha(k/\epsilon)\,b_1(k)\,b_1\left(\frac{k}{\epsilon}\right)
  \right]+\nonumber\\
  &&+
  \left[ 2 \epsilon^3\,\alpha\left(\frac{k}{\epsilon}\right)\,\alpha(k)\,b_1(k) \,b_2(k)+ 2 \alpha(k) \left(\epsilon^3
  \,\alpha(k/\epsilon)\,b_1\left(\frac{k}{\epsilon}\right) b_2(k)+ b_2\left(\frac{k}{\epsilon}\right) \,\alpha(k)\, b_1(k)\right)\right]
\Big\}
  \label{bmsqhhm}
    \,,\\
  B_{hhh}&=&\,
  \frac{\Delta_0^2\,\alpha(k)\,b_1(k)}{k^6}
\left\{ 2 \epsilon^3
\alpha(k/\epsilon)\,b_1\left(\frac{k}{\epsilon}\right) b_2(k)+ b_2\left(\frac{k}{\epsilon}\right) \,\alpha(k)\,b_1(k)
  \right\}
  \,,\label{bmsqhhh}
  \eea
where we keep all contributions. Notice that the structure of the bispectra is very similar to the case of single source,
apart from coefficients depending on $\Pi$, the ratio of the power spectra of the Gaussian fields.
 This implies that when expanded using 
  eqs. (\ref{mb1},\,\ref{mb2}), one finds various contributions depending both on bispectrum ($\fnl$) and trispectrum ($\taunl$ and $\gnl$)
  parameters.
   The qualitative features of the bispectra as a function of the scale remain the same as the ones
   discussed in the single source case. For example, let us
   write the multi-source version of the quantity $C_{fnl}$ that we wrote in eq. (\ref{dcfnl}): it reads
   \be
       C_{fnl}\
    \,=\,\frac{2\fnl\,\left(1+\Pi\right)\,\left(1+2 \epsilon^3\right)\,\alpha\left( k/\epsilon\right)}{
    \alpha(k)\,\left(
 b_g\, \alpha(k) +
  \fnl \beta_2 + \gnl \beta_3 \right)
    }\,.
   \ee
   Also in the multiple source case, this quantity represents a clean probe of the nG parameter $\fnl$ (almost) independent from the value
   of $\gnl$.  
    
   In the next section, we investigate the contributions depending on $\Pi$ that play an important role in defining the properties of the stochastic halo bias.

\section{Stochastic halo bias, and combinations of power spectra and bispectra}
\label{sec:stochastic}

Stochastic halo bias arises when halo overdensities are not fully correlated with matter overdensities
 \cite{Baumann:2012bc,Tseliakhovich:2010kf}. 
 An example is  the inequality
 $$ P_{hh}\,\ge\, b^2 P_{mm}$$
where $b\,=\,P_{mh}/P_{mm}$ is the halo-matter bias. 
 As we
have seen, halo and matter overdensities depend on primordial nG: 
neglecting the effect of shot-noise, the
aforementioned   
stochasticity is associated with inequalities between non-Gaussian parameters. Such inequalities 
 have been well studied in the analysis of nG from inflation, and
are typically 
(but not only \cite{Tasinato:2012js,Byrnes:2011ri}) associated with the presence of multiple sources for the primordial
gravitational potential.
A famous inequality is the 
Suyama-Yamaguchi inequality $\taunl\ge \left({6}/{5}\,\fnl\right)^2$.
   In 
  this section, we will investigate the role of $\gnl$ for characterizing the stochasticity of halo bias, and 
  we will 
   extend the notion of stochasticity to the bispectra. 
   We will make use of 
   some technical results on inequalities among primordial
   non-Gaussian parameters, which we relegate to   
 Appendix \ref{app:ine}. In this section, as in the previous ones, we will not include loop effects.

\subsection{Stochastic  halo bias and power spectra}\label{sec:stocp}

A convenient quantity to quantify the stochasticity of halo bias using power spectra is
 $r_P$, 
  defined as \cite{Baumann:2012bc}
  \be
  r_{P}\,=\,\frac{P_{hh}}{P_{mm}}-\left(\frac{P_{mh}}{P_{mm}} \right)^2\,.
  \ee 
  Using the results of Section \ref{sec:npoint}, we find 
  at large scales $k\to 0$
    \be\label{psin}
  r_P\,=\,\frac{4\,\left( \beta_2 \fnl+3\,\beta_3 \gnl\right)^2\,\Pi}{\alpha_0\,k^4}\,\ge\,0\,.
  \ee
  Hence  $r_P\ge0$, and the equality $r_P=0$ can be obtained when nG
  is absent, or in the single field limit $\Pi\to 0$. Notice that 
   $\fnl$ and $\gnl$ 
   appear in a combination 
   that renders  the identification of 
   their individual effects difficult -- a feature that we already discussed by studying power spectra in the previous sections.

  \smallskip 
  Let us provide a heuristic understanding for this result, extending the arguments of \cite{Baumann:2012bc}
  to include $\gnl$.
     In that paper,
  setting $\gnl=0$, the inequality $r_P\ge 0$ was associated to the Suyama-Yamaguchi inequality $\taunl\ge
  \left(6/5 \fnl\right)^2$. On the other hand, when including $\gnl$, we learn that eq. (\ref{psin})
  reads
  \be\label{psin2}
 \alpha_0\,k^4\, r_P\,=\,\left(\frac{5\,\beta_2}{6}\right)^2\left[ \taunl-\left( \frac{6}{5}\,\fnl\right)^2\right]
 +2\,\beta_2\,\beta_3\,\fnl\,\gnl\,\Pi+\beta_3\,\gnl^2\,\Pi\,\ge\,0\,.
  \ee
  Hence
  we find 
   two additional terms proportional to $\gnl$ besides the first corresponding to the Suyama-Yamaguchi
    inequality.
   
As we have seen in the previous sections,   the halo displacement  can be schematically expressed
as an expansion in terms of local correlation functions of the linearly evolved matter density field 
\be
\delta_h\,=\,b_g \,\delta+\,\beta_2\,[ \delta^2]+\,\beta_3\,[ \delta^3]+\dots
\ee
  Using this expansion, we can schematically express the quantity $r_P$
  as
  \bea
  r_P&=&
  \frac{\langle \delta_h \delta_h\rangle}{\langle \delta \delta\rangle}
  -\frac{\langle \delta_h \delta\rangle^2}{\langle \delta \delta\rangle^2} 
  \nonumber\\
  &=& \beta_2^2\,\left(\frac{\langle [ \delta^2][ \delta^2]  \rangle}{\langle \delta \delta\rangle
  }-
  \frac{\langle [ \delta^2 ] \delta\rangle^2}{\langle \delta \delta\rangle^2}
   \right)+2 \beta_2 \beta_3 \left(
   \frac{\langle [ \delta^3][ \delta^2]  \rangle}{\langle \delta \delta\rangle
  }-
  \frac{\langle [ \delta^2 ] \delta\rangle\,\langle [ \delta^3 ] \delta\rangle}{\langle \delta \delta\rangle^2} 
    \right)+  \beta_3^2 \left( 
    \frac{\langle [ \delta^3][ \delta^3]  \rangle}{\langle \delta \delta\rangle
  }-
  \frac{\langle [ \delta^3 ] \delta\rangle^2}{\langle \delta \delta\rangle^2}
    \right)\label{exprp}\,.
  \eea
  To write the previous formula, we used the schematic notation discussed in Appendix \ref{app:ine} to express
  squeezed and collapsed limits of $n$-pt functions. With $\langle \delta [ \delta^2]\rangle$ we denote the squeezed
  configuration of a 3-pt function, in which one of the momenta is sent to zero: this notation clearly demonstrates
 that a long mode $\delta$ modulates the 2-pt function $\delta [ \delta^2]$.
  With  $\langle[ \delta^2]\, [ \delta^2]\rangle$
  we denote the collapsed configuration of a 4-pt function, in which the momentum connecting the two $[ \delta^2]$
  is going to zero. 
  
The combination in the first term in the RHS of 
equation (\ref{exprp}) relates the collapsed limit of
a four point function with the squeezed limit of the three point function, and
 is associated with the Suyama-Yamaguchi inequality (see eq. (\ref{nongi1b})).
  It corresponds to the first term in (\ref{psin2}). 
 The remaining terms are instead associated to the inequalities 
 (\ref{nongi2}), (\ref{nongi3}), that involve collapsed
 limits of higher point functions, and are at the origin of the additional contributions in (\ref{psin2})
 depending on $\gnl$.  
 The conclusion is that $r_P\ge 0$ does not measure just the SY inequality, but a combination
 of contributions associated with various inequalities: thus, 
 it cannot clearly distinguish between $\fnl$ and $\gnl$.

\subsection{Stochastic  halo bias and the bispectra}\label{sec:stocb}

Stochasticity can be defined also using bispectra: this allows one
to break the degeneracy between $\fnl$ and $\gnl$ also at the level
of stochastic halo bias. 
  Consider the bispectra defined for isosceles triangles, both for the case of single and multiple sources
  as discussed in the previous sections.
   (Squeezed configurations are special cases
in which one of the sides of the triangle has vanishing size, i.e. the parameter $\epsilon$ defined
in  section \ref{sec:isomulti} is large). 
  
  We define the following quantity $r_B$ that uses bispectra to measure stochasticity 
    
  \be
  r_{B}\,=\,\frac{B_{hhh}}{P_{hh}^2}    +\frac{3\,B_{hmm}}{P_{hm}^2} -\frac{3\,B_{hhm}}{P_{hm} P_{hh}} - 
  \frac{B_{mmm}}{P_{hm} P_{mm}}\,.
  \ee
    
  A non-vanishing $r_B$  is associated with stochasticity: indeed, using the same arguments of the previous subsection, one
  can  expand the previous quantity $r_B$ in terms of $\beta_i$, $\fnl$ and $\gnl$ finding a sum of several contributions that
  vanish in the single source limit.
  These contributions, among other things,
    depend on appropriated collapsed limits of six point functions; an example
     of contributions to $r_B$ 
     is the following combination, which
  appears in the expansion for $r_B$
   multiplied by  suitable powers of the available parameters
   
 \be
 {\cal C}\,\equiv\,
 \left[
 \frac{\langle
  [\delta^2]\, [\delta^2] \, [\delta^2]
  \rangle }{\langle [\delta^2]\, [\delta^2] \rangle^2}
  - 
   \frac{\langle
  \delta \,\delta \, \delta 
  \rangle }{\langle \delta \, \delta\rangle 
  \,\langle
  \delta \,   [\delta^2]
  \rangle 
  }- 3 
    \frac{\langle
  \delta \,   [\delta^2]\, [\delta^2]  
  \rangle }{\langle \delta \,  [\delta^2]\rangle 
  \,\langle
   [\delta^2] \,   [\delta^2]
  \rangle 
  }+ 
  3 
    \frac{\langle
  \delta \,   \delta\, [\delta^2]  
  \rangle }{\langle \delta \,  [\delta^2]\rangle^2 
  }\right]\,.
 \ee  
  Using 
   the inequalities among non-Gaussian parameters discussed in  Appendix \ref{app:ine},
    one can check that ${\cal C}$ vanishes in the single source case (neglecting loop
    effects), while it is generally non-vanishing for multiple sources.

  \smallskip
  
  For the specific two-field 
  model  of  section \ref{sec:npoint}, neglecting loop corrections,  one can straightforwardly  check that $r_B$
  is  non-vanishing  only when $\Pi\neq 0$, i.e. in presence of multiple sources.
  We find the following value for $r_B$ at large scales and in the limit of squeezed configurations $\epsilon\gg1$
  \be
  r_B\,=\,\frac{
  \epsilon^2 \fnl \Pi}{2 (\beta_2 \fnl + 3\beta_3 \gnl)} 
  -\frac{
 3\, \epsilon^2\, \fnl\,b_g\,\alpha_0\,k^2\, \Pi}{4 (\beta_2 \fnl + 3\beta_3 \gnl)^2} 
 +{\cal O}(k^4)\,.
  \ee
  Hence, at very large scales $k\to0$, $r_B$ is sensitive to $\fnl$ and its sign:  if $\fnl$ vanishes, then $r_B$ vanishes no matter
  of the size of $\gnl$. Hence $r_B$ can be an useful complementary observable besides $r_P$ to study the stochasticity of halo bias.

  \section{Distinguishing $\gnl$ with Large Scale Structures}
 \label{sec:gnlLSS}
The analytical results of the previous sections indicate that the effects of $\fnl$ and $\gnl$ can be distinguished by measuring bispectra
of halo and matter densities, 
 in a way that is not possible by studying power spectra only.
 Indeed, in the scale-dependent bias of the halo power spectrum the parameters $\fnl$ and $\gnl$ are weighted by
 the same power of the scale $k$, while the bispectra contain different contributions depending on non-Gaussian parameters
  that scale with different powers of $k$, implying it is possible to overcome the degeneracy
   between $\fnl$ and $\gnl$
   without having to study galaxy  $4$-point functions.

 \smallskip
 
 We theoretically analyzed this subject
 in the previous sections, making simplifying assumptions aimed to concentrate on the effects of primordial nG on the halo
 bias.
 While in our theoretical discussions, in order to obtain analytic expressions for our quantities, we focussed on the limit of 
 large scales where   simple approximations  for the transfer functions hold, in this section we 
 numerically investigate the features of the bispectrum also for smaller scales, in order to determine   properties
 of this quantity  that allow  us to distinguish the effects of $\gnl$ from the ones of $\fnl$.
  In particular, we investigate two properties of halo and matter bispectra that we found when analytically studying them in the 
  previous theoretical sections: 1) The dependence on $\gnl$ of the sign of halo bispectra as a function of the scale.
  and 2) The fact that a particular combination of halo and matter bispectra, $C_{fnl}$ (see eq. (\ref{dcfnl})) depends
  on $\fnl$ only, allowing one to cleanly distinguish $\fnl$ from $\gnl$. 
  
 \smallskip
 
To start with, 
  we focus on the single source case and investigate the slope of the halo  bispectrum $B_{hhh}$
 associated with squeezed isosceles triangles in momentum space.
   We consider highly biased halos for all the possible combinations of $\fnl=0, \pm1$ and $\gnl=0,\pm10^{3}, \pm10^{4}$, in the squeezed configuration $\epsilon=100$ (that is, the ratio between 
the long and short sides of an isosceles triangle in momentum space is 100). The aim 
 of our analysis is to confirm the analytical results we determined in the previous sections, and to investigate
at which scales, redshifts and masses a distinct signature of the  effects of $\gnl$ might be observed by future surveys, like EUCLID \cite{Laureijs:2011gra}. On the other hand,
 our analysis intends to concentrate on the effects of primordial nG, without 
  including the non-linear evolution of gravity, nor non-linearities associated with 
   the particular dependence of the mass function on the matter density contast.
  These contributions will be addressed in a future publication.

The starting point of our arguments is eq.\eqref{bshhh} for $B_{hhh}$. In order to obtain the quantities $b_1$, $b_2$ defined in eqs.\eqref{eq:b1}, \eqref{eq:b2}, we compute $\alpha(k,z)$ by using the transfer function from CAMB \cite{CAMB} with the state parameters consistent with the WMAP$5+$BAO$+$SN cosmology \cite{Komatsu:2008hk} (the reason of this choice will be clear later). Then, to evaluate the bias coefficients $b_g$, $\beta_2$, $\beta_3$ defined in eqs.\eqref{eq:bg}, \eqref{eq:beta}, a 
 specific 
mass function must be chosen at the expense of making additional assumptions.


The mass function gives the number density of halos at a given redshift $z$ with mass within $M$ and $M+dM$; under the assumption of universality, the mass function depends only on $\nu=\delta_c/\sigma_M$. It takes  the form \cite{Press:1973iz}
  \be\label{expnh}
  n_h(M,z)=\frac{\overline{\rho}_m}{M^2} f\left(\nu\right)\biggr\vert\frac{d\ln\sigma^{-1}_M}{d\ln M}\biggr\vert,
  \ee
where the variance $\sigma_M$ of the smoothed linear density contrast $\delta_M$ is given by
  \be
  \sigma_M(z)^2\equiv\int\,\frac{d\vec{k}}{(2\pi)^3}W_M^2(k)\bar{\alpha}^2(k,z)P_\Phi(k)
  \ee
and we replace the spherical collapse threshold $\delta_c=1.686$ with $1.42$, to improve
  the agreement between the barrier model and simulations  \cite{Pillepich:2008ka, Grossi:2009an}. When
   the primordial gravitational potential is exactly
   Gaussian, the
    mass function $f$ appearing in (\ref{expnh}) assumes the 
    Press-Schechter form
  \be\label{psdis}
  f_{PS}(\nu)=\sqrt{\frac{2}{\pi}}\,\nu\,e^{-\frac{\nu^2}{2}}.
  \ee
  
  For non-Gaussian  primordial fluctuations, 
$f_{PS}$ must be replaced by a more accurate mass function. 
Several mass functions that account for non-Gaussian initial conditions have been proposed in the literature
 (see for example \cite{Matarrese:2000iz,LoVerde:2007ri} and the reviews \cite{Desjacques:2010jw,Verde:2010wp}).
 For our analysis, we will
 adopt the results of
   Smith, Ferraro, and LoVerde \cite{Smith:2011ub}, based on the Edgeworth expansion;
   these
 provide  a useful, theoretically motivated expression for halo bias which agrees with $N$-body simulations. 

The Edgeworth expansion can be applied within  the barrier crossing model by building up the PDF as a series of higher-order cumulants (defined as $\kappa_n\,=\,\langle \delta_M^n \rangle_c/\sigma_M^n$) times the Press-Schechter Gaussian distribution
(\ref{psdis}). In order to have a positive definite distribution, it is important to specify the order of the cumulant
 at which the series is truncated  (see \cite{LoVerde:2007ri} for further details, and also \cite{Desjacques:2009jb,Maggiore:2009hp} for additional papers that include
 the effects of kurtosis).
 
  Fortunately,  for the weakly nG regime we are interested in, the Edgeworth expansion converges rapidly, providing a useful approximation as long as $1\gg\kappa_3\gg\dots\gg\kappa_n$. Within this regime, the Edgeworth expansion corresponds simply the Press-Schechter mass function plus first-order corrections in $\fnl$ and $\gnl$:

\bea
  \label{eq:edgeworth}
n(M)=\frac{2\rho_0}{M}\,\biggr\vert\frac{d\ln\sigma^{-1}_M}{d M}\biggr\vert\, \frac{e^{-\nu^2/2}}{(2\pi)^{1/2}}\Biggr[\nu &+& \fnl \left(\kappa^{(1)}_{3}(M)\frac{\nu H_3(\nu)}{6}-\frac{d\kappa_3^{(1)}/dM}{d(\ln\sigma^{-1}_M)/dM} \frac{H_2(\nu)}{6}\right)+\nonumber \\ &+& \gnl \left(\kappa^{(1)}_{4}(M)\frac{\nu H_4(\nu)}{24}-\frac{d\kappa_4^{(1)}/dM}{d(\ln\sigma^{-1}_M)/dM} \frac{H_3(\nu)}{24}\right) \Biggr],
\eea
where $H_n$ is $n$-th Hermite polynomial and we defined the $n$-th nG cumulant

\bea
\kappa_3(M)&=\fnl \kappa_3^{(1)}(M)\approx\fnl\left(6.6\cdot10^{-4}\,\right)\left[1-0.016\ln\left(\frac{M}{h^{-1}M_{\odot}}\right)\right]\label{eq:kappa_3}\,,\\
\kappa_4(M)&=\gnl \kappa_4^{(1)}(M)\approx\gnl\left(1.6\cdot10^{-7}\,\right)\left[1-0.021\ln\left(\frac{M}{h^{-1}M_{\odot}}\right)\right]\label{eq:kappa_4}.
\eea
The approximated results above are taken from \cite{LoVerde:2011iz}.

\smallskip
{ It is important to underline that the Edgeworth expansion  is not in a universal form, as an explicit dependence on $M$ appears. However, the cumulants $\kappa_3$ and $\kappa_4$ are weakly dependent on it (see eqs.\,\eqref{eq:kappa_3},\,\eqref{eq:kappa_4}), so that we are justified to drop $M$ and assume them as constants. Although this is not true in general for $[d\kappa_3^{(1)}/dM]/[d(\ln\sigma^{-1}_M)/dM]$ and $[d\kappa_4^{(1)}/dM]/[d(\ln\sigma^{-1}_M)/dM]$, we have checked it to be a reasonable approximation within the mass range we are interested in.}

Under these approximations universality is recovered, and within the validity of the Edgeworth expansion, the bias coefficients are obtained by taking the derivatives of eq.\eqref{eq:edgeworth}.
Decomposing
\be
b_g\,=\,b_{g\,0}+\fnl b_{\fnl\,0}+\gnl b_{\gnl\,0}\,,
\ee
we have
\bea 
b_{g\,0}&=&\,1 + \frac{\nu^2 - 1}{\delta_c}\label{eq:bg0}\,, \\
b_{\fnl\,0}&=&\,-\kappa^{(1)}_{3}(M)\left(\frac{\nu^3 - \nu}{2\delta_c}\right)-\frac{d\kappa_3^{(1)}/dM}{d(\ln\sigma^{-1}_M)/dM} \left(\frac{\nu + \nu^{-1}}{6\delta_c}\right) \,,\\
b_{\gnl\,0}&=&\,-\kappa^{(1)}_{4}(M)\left(\frac{\nu^4 - 3\nu^2}{6\delta_c}\right)-\frac{d\kappa_4^{(1)}/dM}{d(\ln\sigma^{-1}_M)/dM} \left(\frac{\nu^2}{12\delta_c}\right) \,.
\eea
Moreover,
\bea
\beta_2&=&\,2\nu^2 - 2 \label{eq:beta2}\,,\\
\beta_3&=&\,\kappa^{(1)}_{3}(M)\left(\frac{\nu^3 - \nu}{2}\right)-\frac{d\kappa_3^{(1)}/dM}{d(\ln\sigma^{-1}_M)/dM} \left(\frac{\nu - \nu^{-1}}{2}\right)\label{eq:beta3}.
\eea
The physical interpretation of the results above is the following: $b_{g0}$ is the usual Gaussian bias in Eulerian coordinates, while the terms $b_{\fnl\,0}$ and $b_{\gnl\,0}$ describe the scale-independent shift to $b_{g\,0}$.
 Unfortunately they cannot be used to constrain primordial NG, since the bias of a real tracer population is {not known a priori}. { Then, $\beta_2$ is the well-known scale-dependent bias for an $\fnl$ cosmology. By neglecting the explicit mass dependence of the Edgeworth expansion, we have forced eq.\eqref{eq:beta2} to satisfy the relation $\beta_2=2\delta_c(b_{g\,0}-1)$ found in \cite{Slosar:2008hx}; this is valid for a universal mass function}. Finally, $\beta_3$ describes the scale-dependent bias introduced by the $\gnl$ parameter.


\smallskip

 Recall that the mass function here is initially computed  in Lagrangian coordinates, therefore the $+1$ in eq.\eqref{eq:bg0} takes into account the dynamical effects produced by linear  gravity and bring us to Eulerian coordinates \cite{Mo:1995cs}.  Let us
 emphasize again that here we are only including the linear effects of gravity. Non-linear effects would 
 add further non-Gaussian features to bispectra -- not specifically due uniquely to primordial nG, the focus of this paper   -- that we intend to analyze in a future publication. 

\smallskip

Smith et al. compared the Edgeworth prediction for $\beta_3$ with $N$-body simulations, finding that it breaks down at {lower} halo mass, $M\lesssim10^{14}h^{-1}M_{\odot}$. To fix this problem, they first noted that eq.\eqref{eq:beta3} can be written as
\be
\beta_3=\,\kappa_3\left[-1+\frac{3}{2}\left(\nu -1\right)^2 + \frac{1}{2}\left(\nu - 1\right)^3\right] - \frac{d\kappa_3}{d\ln\sigma^{-1}}\left(\frac{\nu-\nu^{-1}}{2}\right),
\ee
then, they found a good agreement with simulations by changing the coefficients of the polynomial in the brackets as follows
\be
\label{eq:beta3ast}
\beta_3^{\ast}=\,\kappa_3\left[-0.7+1.4\left(\nu -1\right)^2 + 0.6\left(\nu - 1\right)^3\right] - \frac{d\kappa_3}{d\ln\sigma^{-1}}\left(\frac{\nu-\nu^{-1}}{2}\right).
\ee
Our numerical analysis can be easily performed by using the following fitting functions \cite{Smith:2011ub}
\bea
\kappa_3&=&\,0.000329\left(1+0.09 z\right) b_{g\,0}^{-0.09}\label{eq:kappa_32}\,,\\
\frac{d\kappa_3}{d\ln\sigma^{-1}_M}&=&\,-0.000061\left(1+0.22 z\right) b_{g\,0}^{-0.25}\label{eq:dkappa_3}.
\eea
The quantities \eqref{eq:bg0}-\eqref{eq:beta2},\eqref{eq:beta3ast} provide the necessary terms to describe highly biased tracers. 
All the results provided by Smith et al. have been tested against $4$ simulations with Gaussian initial conditions, $5$ simulations with $\fnl=\pm250$ and $3$ simulations with $\gnl=\pm2\cdot 10^{6}$, for a total of $20$ simulations in the WMAP$5+$BAO$+$SN cosmology. 
 Given that these results have been well tested on several simulations, 
we assume the same dataset, and  we do not expect to see any radical change in our order-of-magnitude analysis for a slightly different $\Lambda$CDM cosmology. However, further simulations should be carried out to test the validity of $\beta_3^{\ast}$ for the low values of $\fnl$ and $\gnl$ we are interested in.

\smallskip

Let us now present our numerical results and their interpretation.
 Fig. 1 
 displays the absolute value of the halo bispectrum (using a $\log$-scaling), for
 halos of mass $M=10^{13}M_{\odot}$, at redshift $z=1$, and $k$-space triangles with squeezing parameter $\epsilon=100$.
  Each of the three plots in this figure 
  shows the results for a different value of $\fnl$: $\fnl=-1$ for 
Case 1,  $\fnl=0$ for Case  2,  and $\fnl=1$ for
Case 3. 
 The lines on each plot correspond to different values of $\gnl$. Each point on the $k$-axis actually involves three
different values of the wavenumber specifying the three sides of a squeezed triangle in momentum space:
 two at $k_1=k_2=k$ (the long equal sides) and one at $k_3=k/\epsilon$ (the small side). The plots span an interval
  of $k$ between $10^{-2}$
 and $10^{-1}\,h/\text{Mpc}$, corresponding to 
   scales within ranges
 that will be probed by future observations made by EUCLID, under the hypothesis that this survey will gain one order of magnitude
 in scale 
 with respect to BOSS \cite{Ross:2012sx, Anderson:2012sa}.  
Since negative values of the bispectrum are possible, we plot the absolute value of $B_{hhh}$. The presence of cusps indicate the changing from positive to negative values (or vice-versa). For the sake of clarity, we use dashed lines for negative values of the bispectrum and solid lines for positive ones.

 \begin{figure}[H]\label{fig:first}
\begin{center}
\subfloat{\includegraphics[width=.5\textwidth]{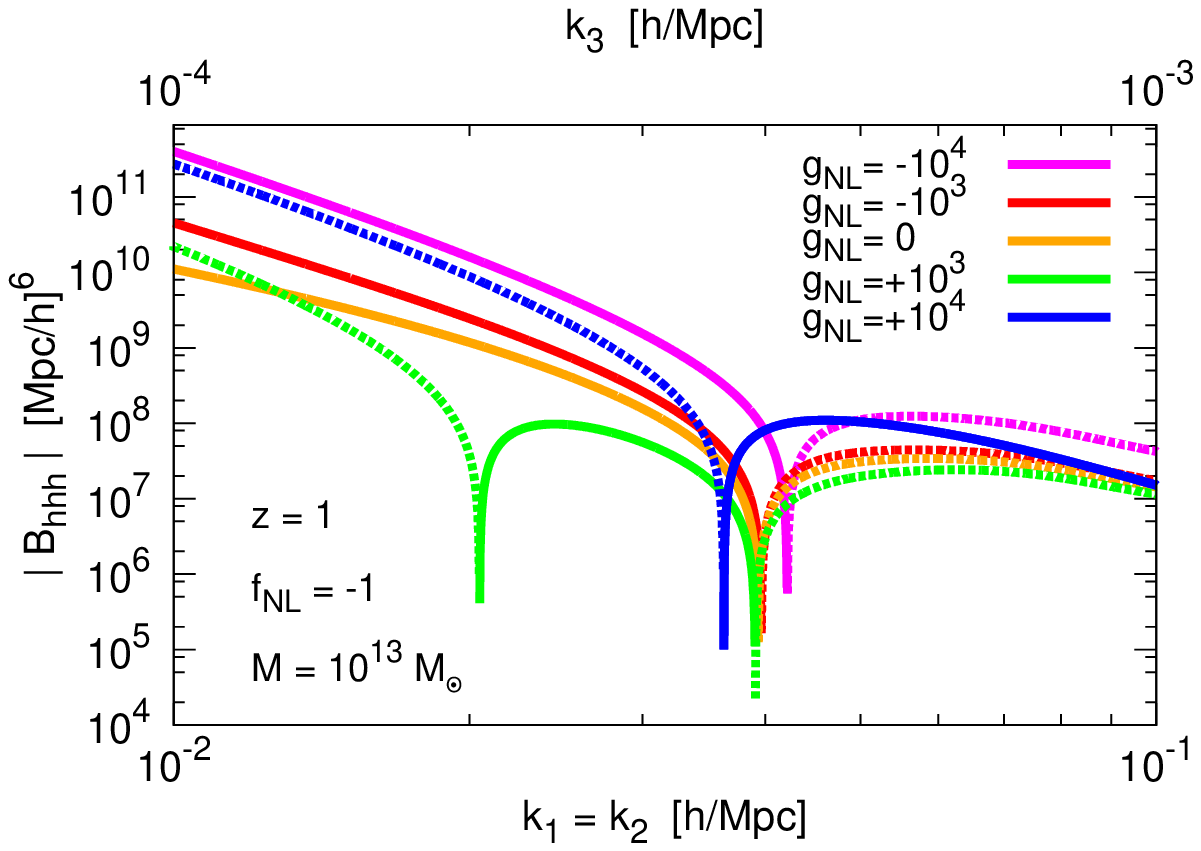}} \\
\subfloat{\includegraphics[width=.5\textwidth]{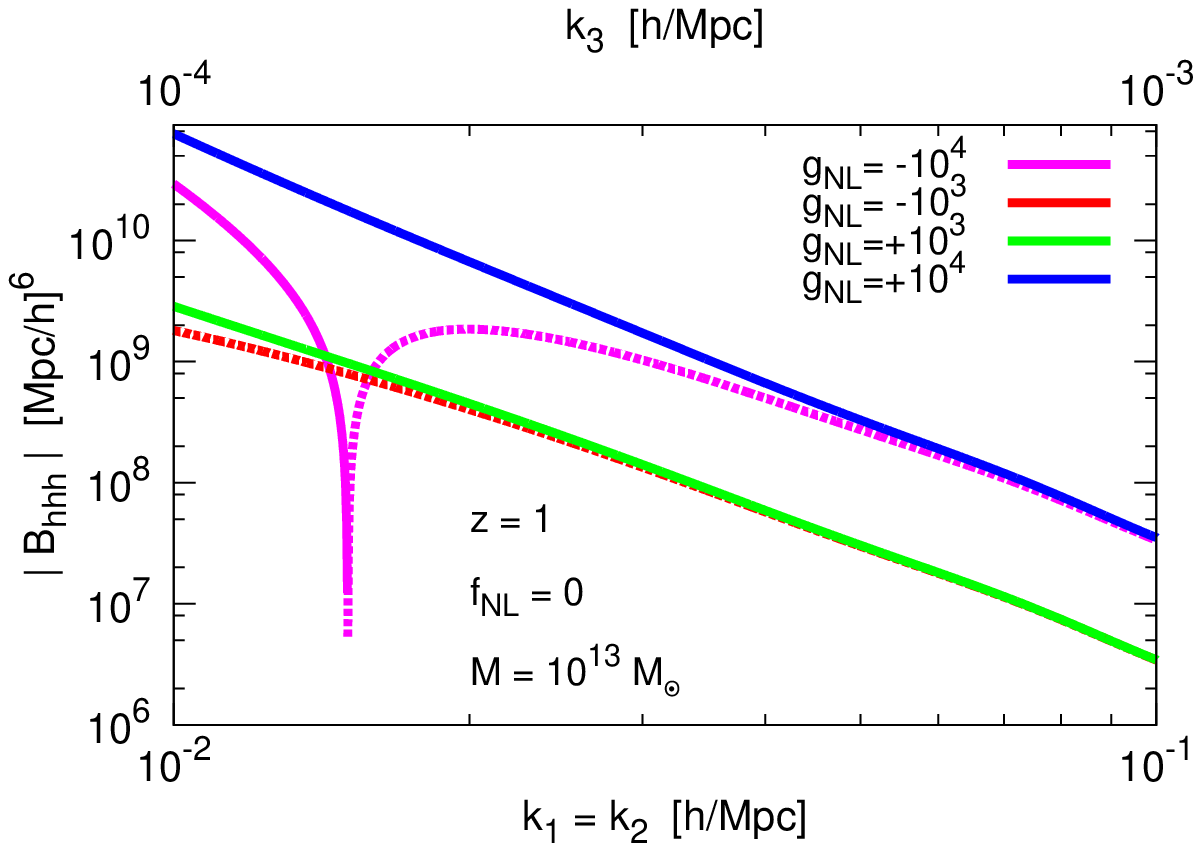}} 
\subfloat{\includegraphics[width=.5\textwidth]{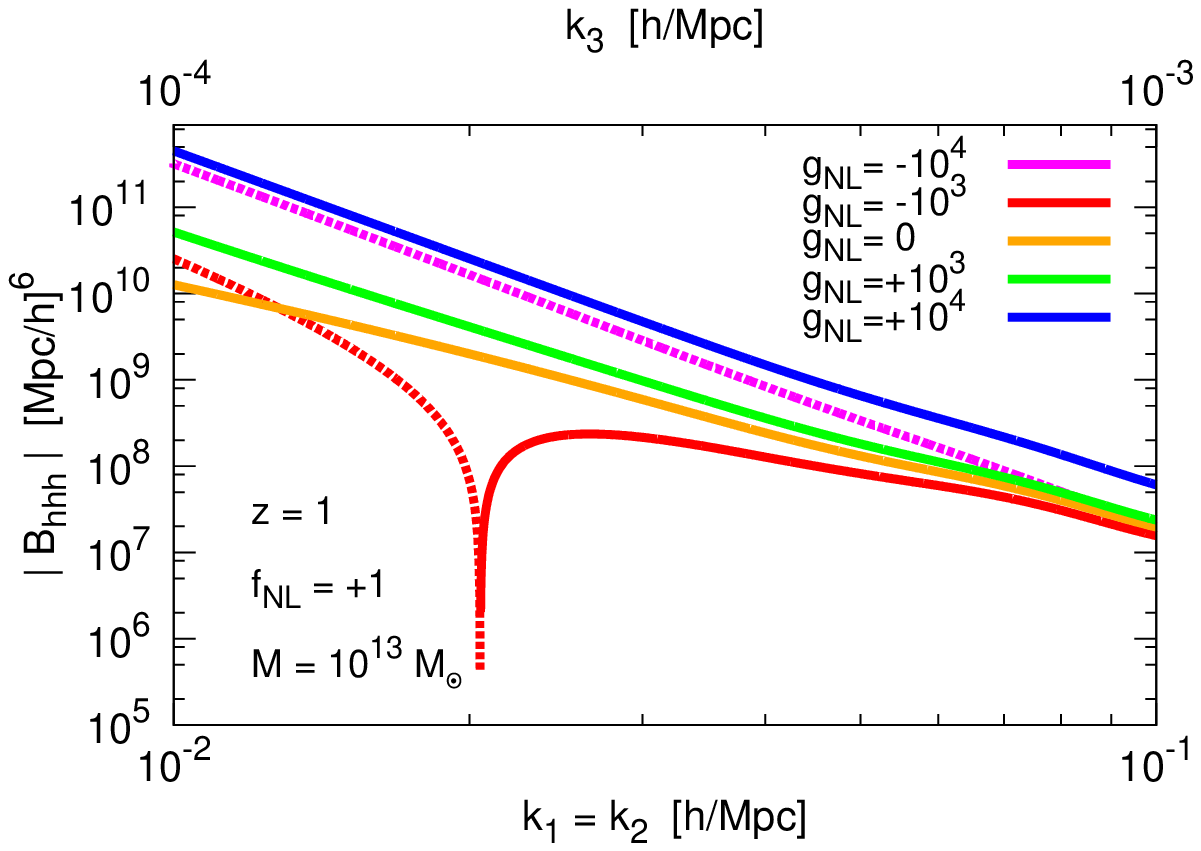}} 
\end{center}
\begin{caption}\\ \emph{Top}: The absolute value of the halo bispectrum $B_{hhh}$ in the squeezed configuration $k_1=k_2=100 k_3$, for halo mass $M=10^{13}M_{\odot}$, redshift $z=1$, $\fnl=-1$. Different lines correspond to the values  $\gnl=0, \pm 10^{3}, \pm 10^{4}$. We use solid lines to describe positive values and dotted lines for neagative values; the presence of a cusp indicates a change of sign. The pattern of zeros in the plotted range of wavenumbers can be used to distinguish $\fnl$ from $\gnl$: see the text for further details. \\ \emph{Bottom left}: Same as the top panel, but for $\fnl=0$. For $\gnl=-10^{4}$ a zero is present. \\ \emph{Bottom right}: Same as top panel, but for $\fnl=1$. Notice that a single zero appears, for $\gnl=-10^{3}$.
\end{caption}
\end{figure}
 An  
 inspection of these plots suggests an important qualitative feature of these results. 
 %
  If the parameters are contained within appropriate intervals, the halo bispectrum as 
  a function of $k$  changes sign more times if   $\gnl$ has opposite sign with respect to $\fnl$. 
  For example, in 
  Fig. 1, Case 1
 corresponding to $\fnl=-1$, we see that the green curve, corresponding to the logarithm 
  of the bispectrum for $\gnl=10^3$, changes sign twice instead of once as the other curves. Moreover, the positions 
   of the zero at smaller scales,  that exists for all the curves, depends on the value of $\gnl$.
    In  Fig. 1, Case 2 with $\fnl=0$, the bispectra for different values of $\gnl$ never change sign 
   in the interval of scales we consider. Finally, in 
  Fig. 1, Case 3 with $\fnl=+1$, 
  the red curve with $\gnl\,=\,-10^3$ changes sign while all the other curves do not.

  The qualitative features in the bispectra profiles as a function of scale 
  become even more pronounced when considering higher redshifts or higher
  mass objects. 
To show this behaviour, we present
 plots displaying 
 the halo bispectrum for two additional cases. The first set, displayed in Fig. 2, is for $M=10^{13}M_{\odot}$, $z=2$, $\epsilon=100$,
  while the second one is for $M=10^{14}M_{\odot}$, $z=1$, $\epsilon=100$ and is displayed in
   Fig. 3. 
Each plot uses the same conventions as Fig. 1.
The qualitative considerations we
made above 
 are still valid. In particular, we see, at higher mass or redshift than our fiducial choice ($M=10^{13}M_{\odot}$, $z=1$, $\epsilon=100$), an increase in the absolute value of the halo bispectrum. Interestingly, the various contributions to the bispectrum weighted by different powers of $k$ make the magenta curve for $\gnl=-10^{4}$ in the plots with $\fnl=+1$ change sign, in both cases presented here. This new feature is due to the
dependence of 
 the bias parameters $b_1$, $b_2$ on mass and redshift. 
 \begin{figure}[H]
\centering
\subfloat{\includegraphics[width=.5\textwidth]{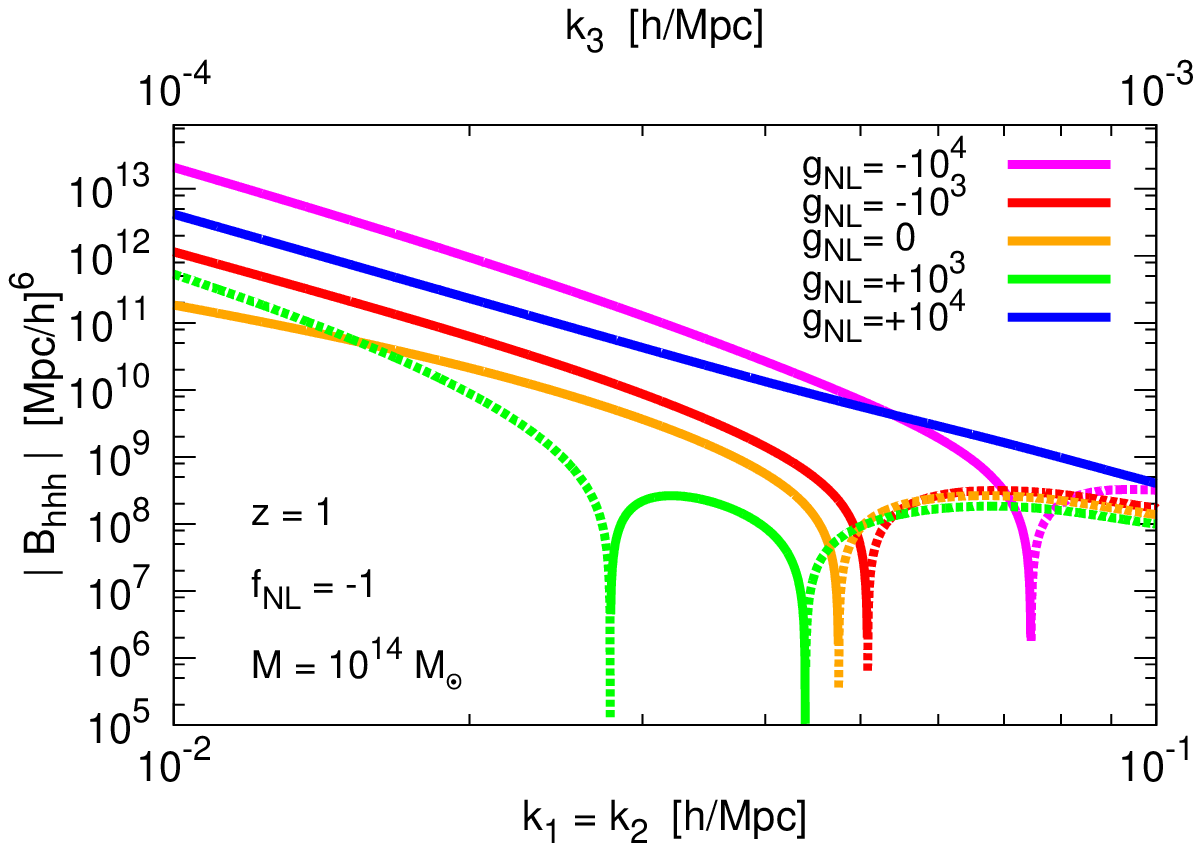}} 
\label{fig:first2}
\subfloat{\includegraphics[width=.5\textwidth]{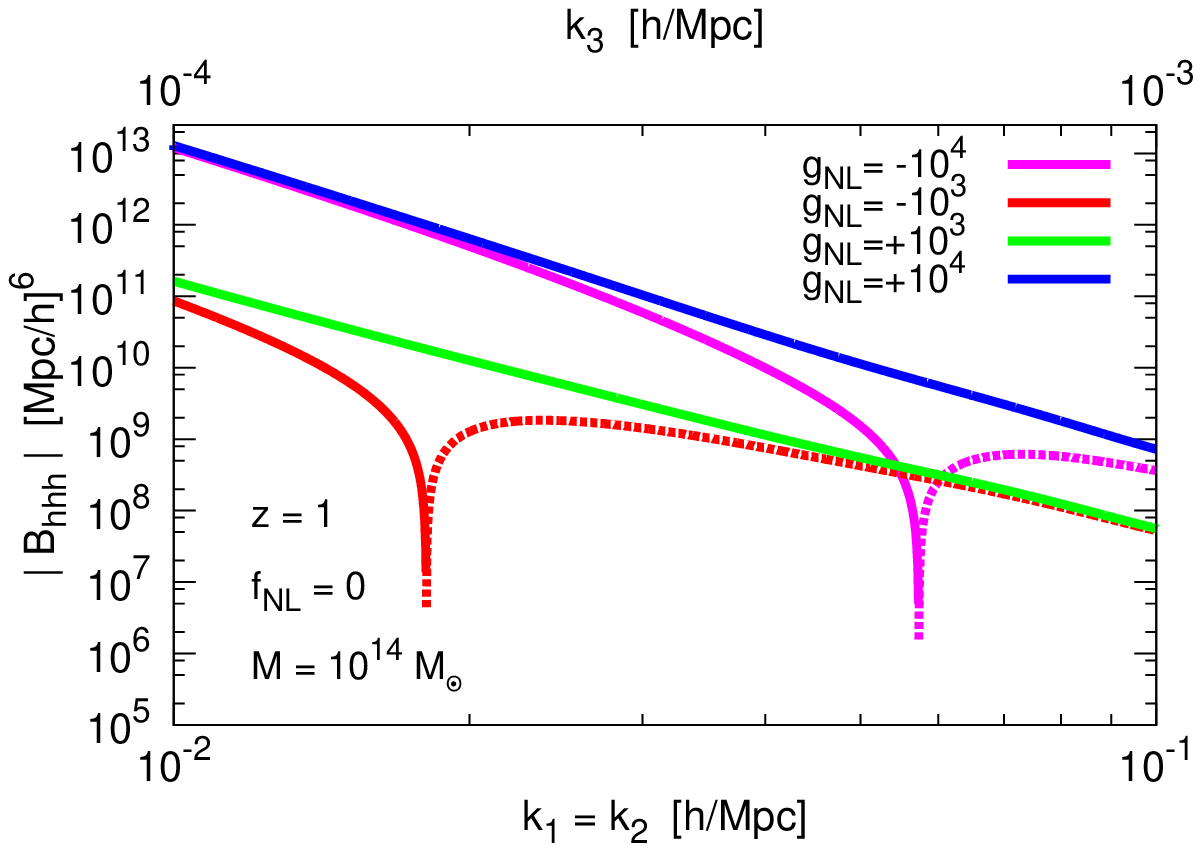}} 
\subfloat{\includegraphics[width=.5\textwidth]{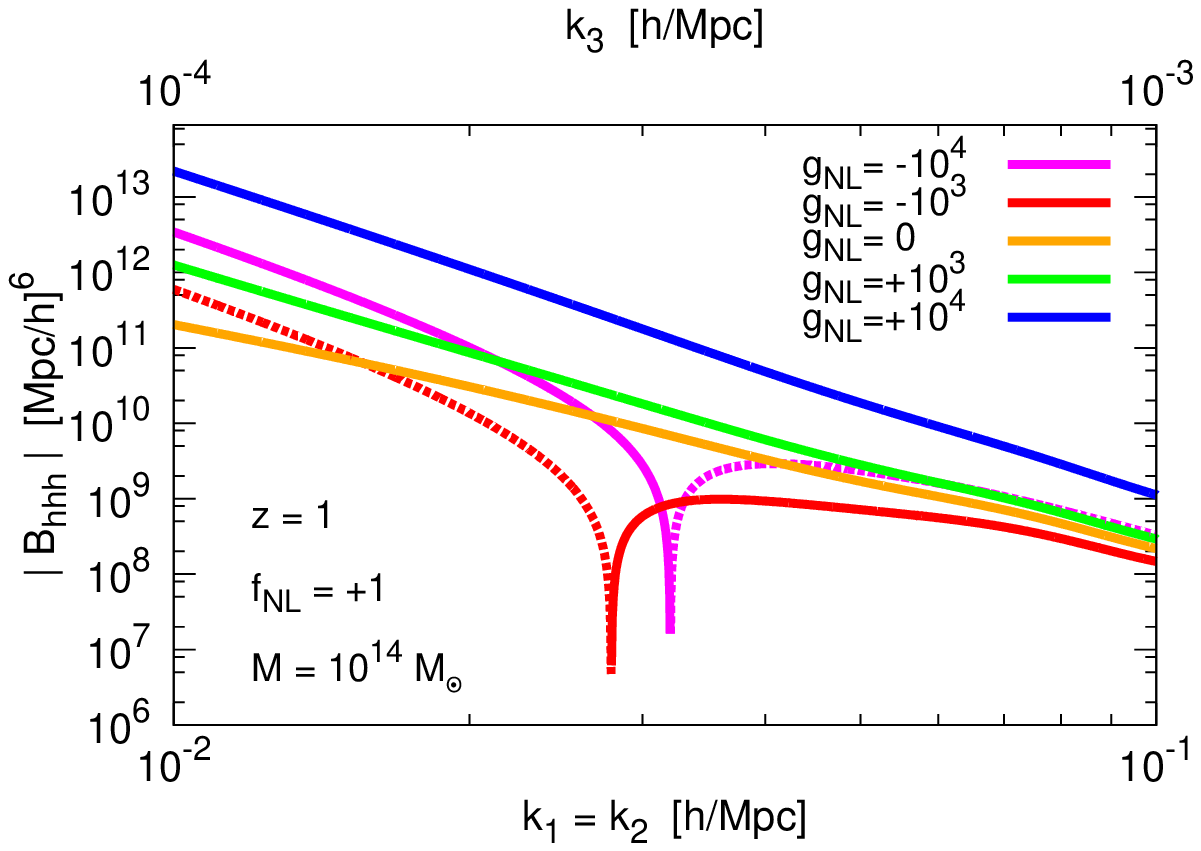}} 
\begin{caption}\\
 Same as Fig. 1, but for $M = 10^{14}M_{\odot}$. The amplitude of the bispectra have increased considerably over the case with $M = 10^{13} M_{\odot}$ (shown in Fig. 1), the positions of the zeros have changed for $f_{NL} = -1$ (top panel, and the number of zeros has increased for $f_{NL} = 0, +1$ (bottom right panel).
\end{caption}
\end{figure}
 \begin{figure}[H]
\centering
\subfloat{\includegraphics[width=.5\textwidth]{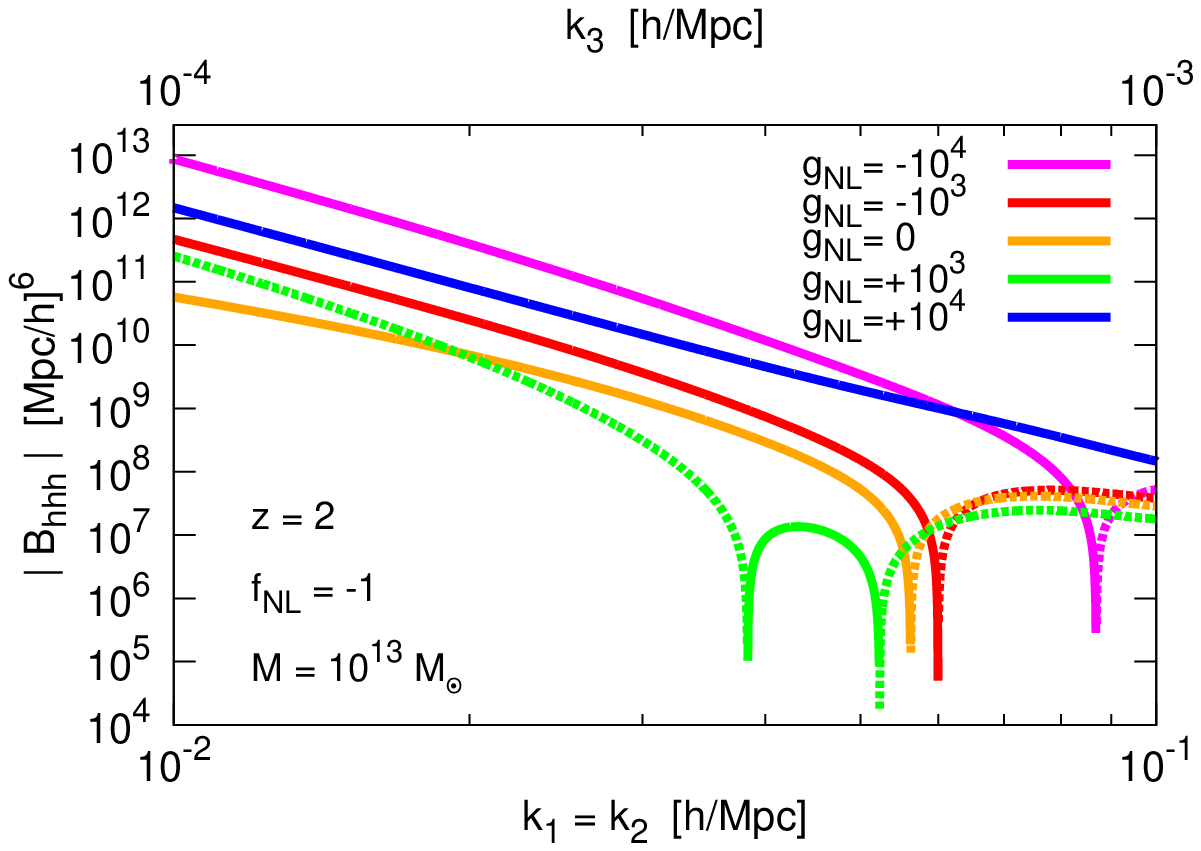}} 
\label{fig:first3}
\subfloat{\includegraphics[width=.5\textwidth]{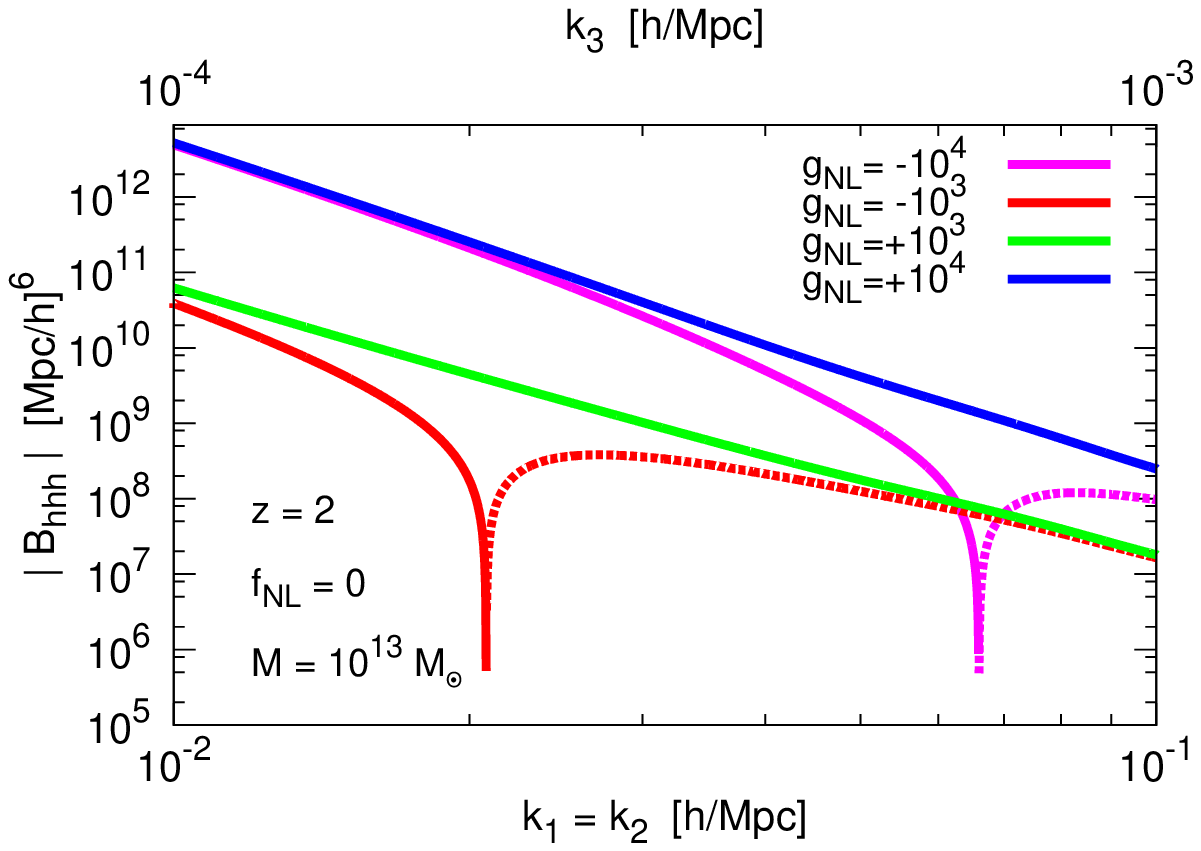}} 
\subfloat{\includegraphics[width=.5\textwidth]{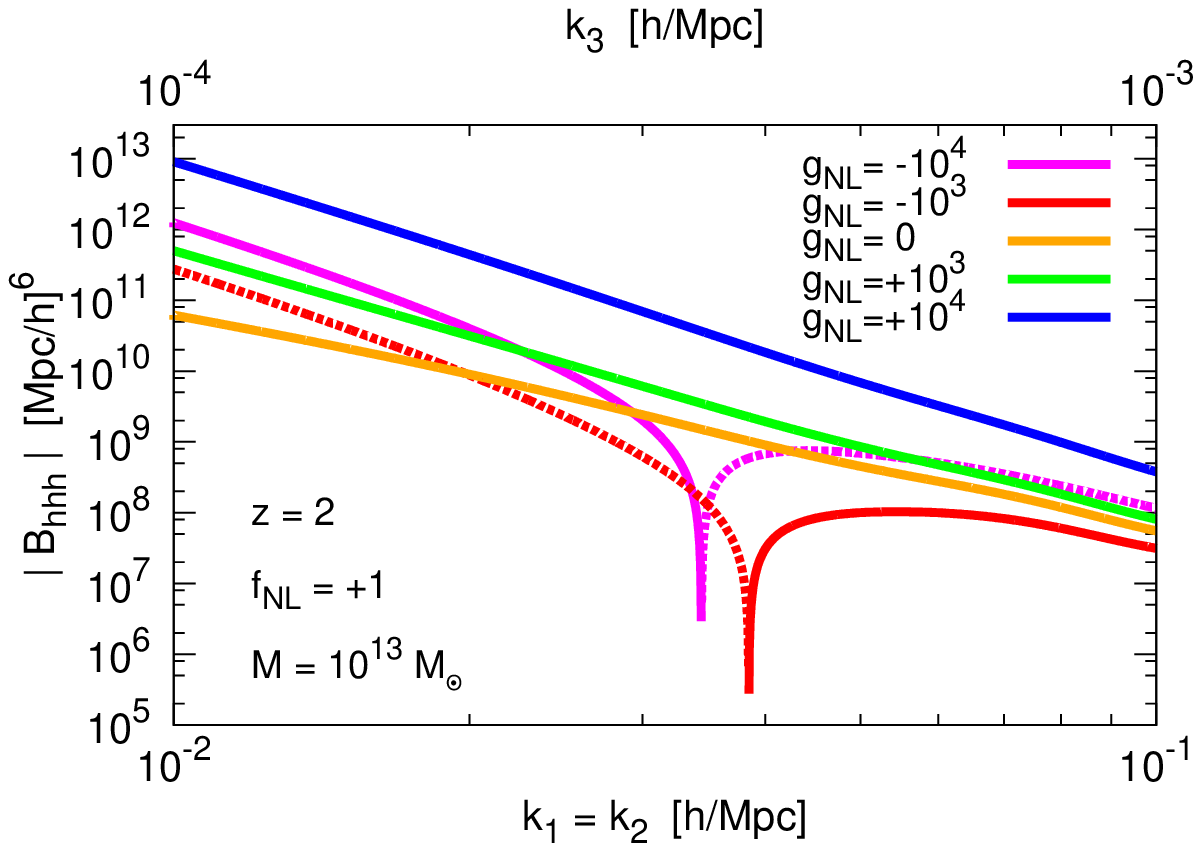}} 
\begin{caption}\\
 Same as Fig. 1, but for $z=2$. The amplitude of the bispectra have increased significantly over the case of $z=1$ (shown in Fig. 1) and the positions of the zeros differ from those observed both in Fig. 1 and Fig. 2.
\end{caption}
\end{figure}
  Hence, the previous plots 
  show that the presence or absence of $\gnl$  
  affects the bispectra profiles: in particular it
  governs  how many times the halo bispectrum   changes sign 
  as a function of the scale.
  We already discussed this qualitative behavior in Section \ref{subsec:squeezed},  when analytically studying
  the roots of the bispectrum in the large scale limit. The numerical results of this section confirm those analytical findings, 
  indicating that the qualitative 
   profile of the halo bispectrum  for isosceles triangles 
       as a function of the scale can provide an interesting signature
  of the presence of $\gnl$.  Hence, we learn that the fact that the bispectrum changes sign in multiple locations as a function 
  of the scale is a distinctive signal of the presence of  $\gnl$.
    Optimistically,   
    this
  can be used to constraint values of $\gnl$ smaller 
    than the ones that can be tested by CMB.
     These features  might be observed in cases in which large values of $b_g$ are realized, since in this case 
    the bispectrum generated by primordial nG is more significant. Thus, the bispectrum is greater for larger halo masses $M$ and, at constant $M$, greater at higher redshift.    For example,
    in Fig. 4
we provide an example of halos at redshift $z=2$, for $\fnl=-1$. In this case, we see that the green curve corresponding to
 the halo bispectrum changes
 sign twice for  
  $\gnl=400$. 
   An observation of this phenomenon can then   probe quite small values
  for $\gnl$.   
    We can also compare the  results of    Fig. 4 with  the analytical formulae for the zeros of the bispectrum,
  eqs (\ref{kroot1}) and (\ref{kroot2}).
For the halo mass, redshift and $\epsilon$ parameter we are considering, 
we find
$
b_g\,\simeq\,4.3$,
$\beta_2\,\simeq\,9.5$, 
$\beta_3\,\simeq\,1.1\,\times\,10^{-3}
$.
Computing the positions of the zeros of the bispectrum for the case $\gnl=400$ with the analytical formulae
of eqs (\ref{kroot1}) and (\ref{kroot2}), one finds the zeros at $k_{root}^{(1)}\simeq1.2 \cdot 10^{-2}$ and $k_{root}^{(2)}\simeq4.9 \cdot 10^{-2}$ h/Mpc, in agreement with the numerical results within one order of magnitude. This implies that our analytical findings, although obtained in the approximation in which we neglect the scale dependence of the transfer function, provide reasonably accurate predictions
 for the positions of the zeroes. 
\begin{figure}[H]
\centering
\subfloat{\includegraphics[width=.6\textwidth]{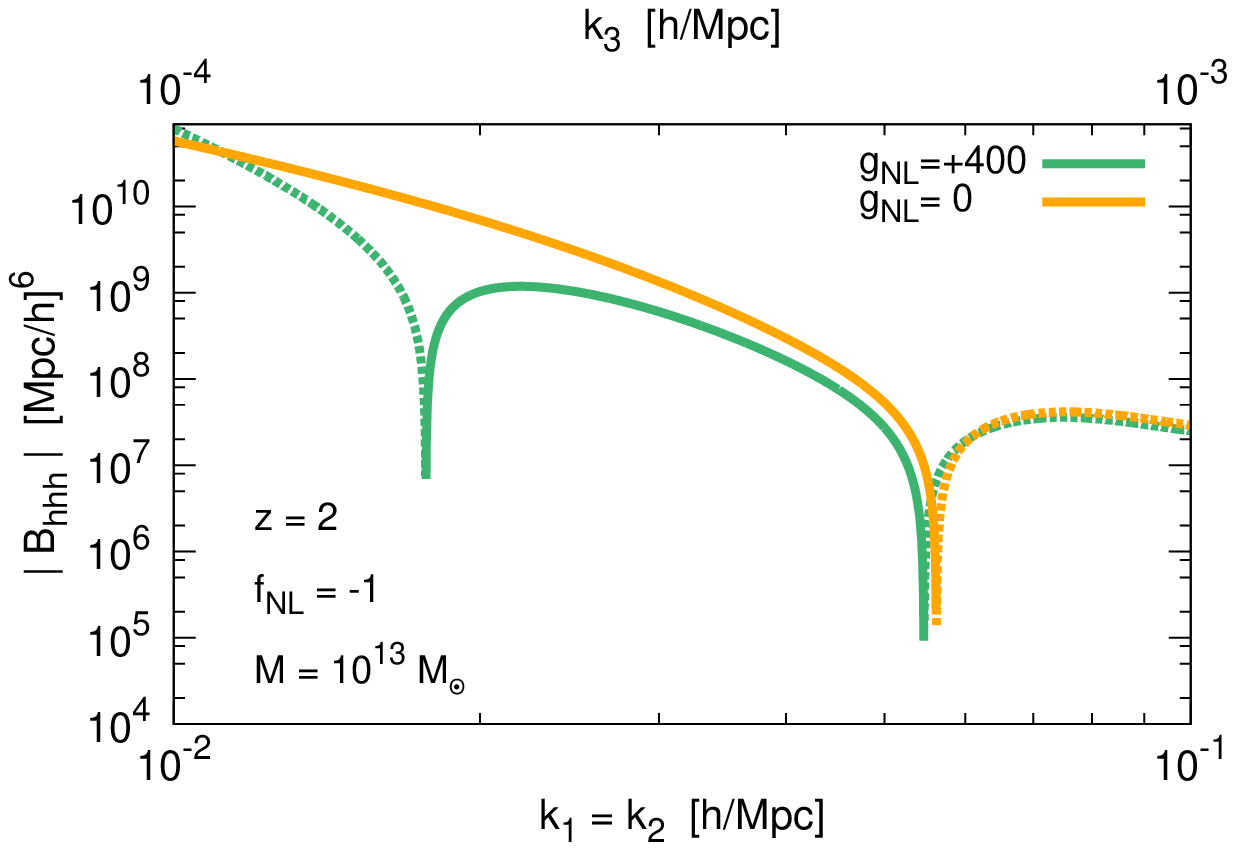}} 
\begin{caption}\\
 The plot shows the absolute value of the halo bispectrum $B_{hhh}$ in the squeezed configuration $k_1=k_2=100 k_3$, for halo mass $M=10^{13}M_{\odot}$ and redshift $z=2$.  $\fnl$ is kept constant and equal to $-1$, while different lines correspond to the values  $\gnl=0, 400$. We use solid lines to describe positive values and dotted lines for negative values; the presence of a cusp indicate a change of sign. $\fnl=-1$ produces a specific pattern of zeros in the line with $\gnl=400$. If this feature can be observed in the halo bispectrum, it will improve the constrain on $\gnl$ to few hundreds.
\end{caption}
\label{fig:fourth}
\end{figure}
We now present a plot that represents  the second theoretical observation  we made in the previous sections: if it 
 is  possible to probe and observe 
  bispectra correlating  halo and matter densities (for example using
gravitational lensing) then by analyzing combinations such as  $C_{fnl}$ of eq. (\ref{dcfnl}) we can have clean measurements of $\fnl$
only, with a negligible contamination of $\gnl$. In  Fig. 5 we represent this quantity for different values of $\fnl$, halo masses and
redshifts; the lines are practically insensitive on $\gnl$, and depend only on $\fnl$ and on the halo mass. 
 This feature, as we discussed around eq. (\ref{dcfnl}), is valid at all scales (although at smaller scales one should take
 into more proper account the non-linear effects of gravity). 
This concretely shows that an observation of a  non-vanishing value for the quantity $C_{fnl}$ would be a clean indication of a non-vanishing 
$\fnl$. Combined with other measurements (for example associated with power spectra) this fact could then be used to obtain
independent constraints on $\gnl$. 
\begin{figure}[H]
\centering
\subfloat{\includegraphics[width=.5\textwidth]{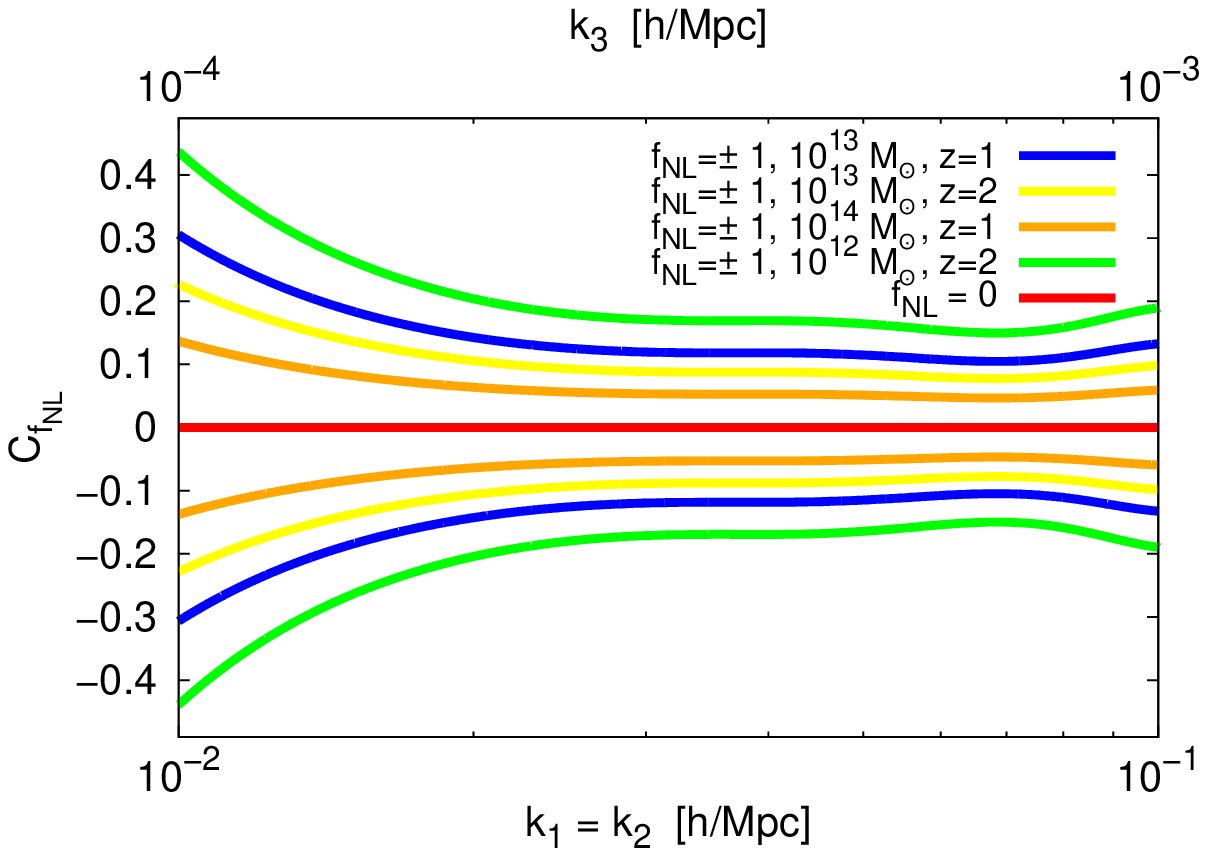}} 
\begin{caption}\\The plot shows the quantity $C_{fnl}$ in the squeezed configuration $k_1=k_2=100 k_3$, for different halo masses ($M=10^{12}M_{\odot}, 10^{13}M_{\odot}, 10^{14}M_{\odot}$) and redshifts ($z=1,2$). $C_{fnl}$ is positive for $\fnl=+1$ and negative for $\fnl=-1$. Remarkably, the lines are insensitive to the values of $\gnl$.
\end{caption}
\label{fig:five}
\end{figure}
We conclude that  the scale dependence of bispectra 
of isosceles configurations in momentum space have
 qualitative features  that might allow us to distinguish the effects of different nG parameters, in a way that is not possible by 
 studying power spectra only.

\smallskip

 While in the previous discussion we neglected the effects of
  non-linear  gravity
   in the results for the bispectra,
    let us end this section with a preliminary 
  analysis of its contributions. 
   Focussing on squeezed 
  configurations, 
  the next plot shows a comparison between the amplitude of bispectra
   associated with
    primordial nG only (color lines) and bispectra due uniquely to the non-linear effects of gravity
  (black lines).   We divide the bispectrum due to gravity in  two contributions. 
   The black dotted line corresponds to the amplitude of bispectra associated only with
  linear bias $b_{g\,0}$, and is controlled by the so-called $F_2$ function.
     This corresponds to the three point function
  for first order fluctuations. 
   The black continuous line, instead, is the contribution associated
  with the quadratic coefficient  \footnote{
  To avoid confusion with eq.\eqref{eq:b2} we call it here and in Appendix \ref{app:comparison} as $B_2$, although in the literature it is usually referred to as $b_2$.} $B_2$ of the local nonlinear biasing model of \cite{Fry:1992vr}. We make the choice $B_2\,=\,2\,b_{g\,0}$ in order to be more conservative respect to \cite{Jeong:2009vd}
   (See 
the review    \cite{Bernardeau:2001qr} for more details.)

  \begin{figure}[H]
\centering
\subfloat{\includegraphics[width=.5\textwidth]{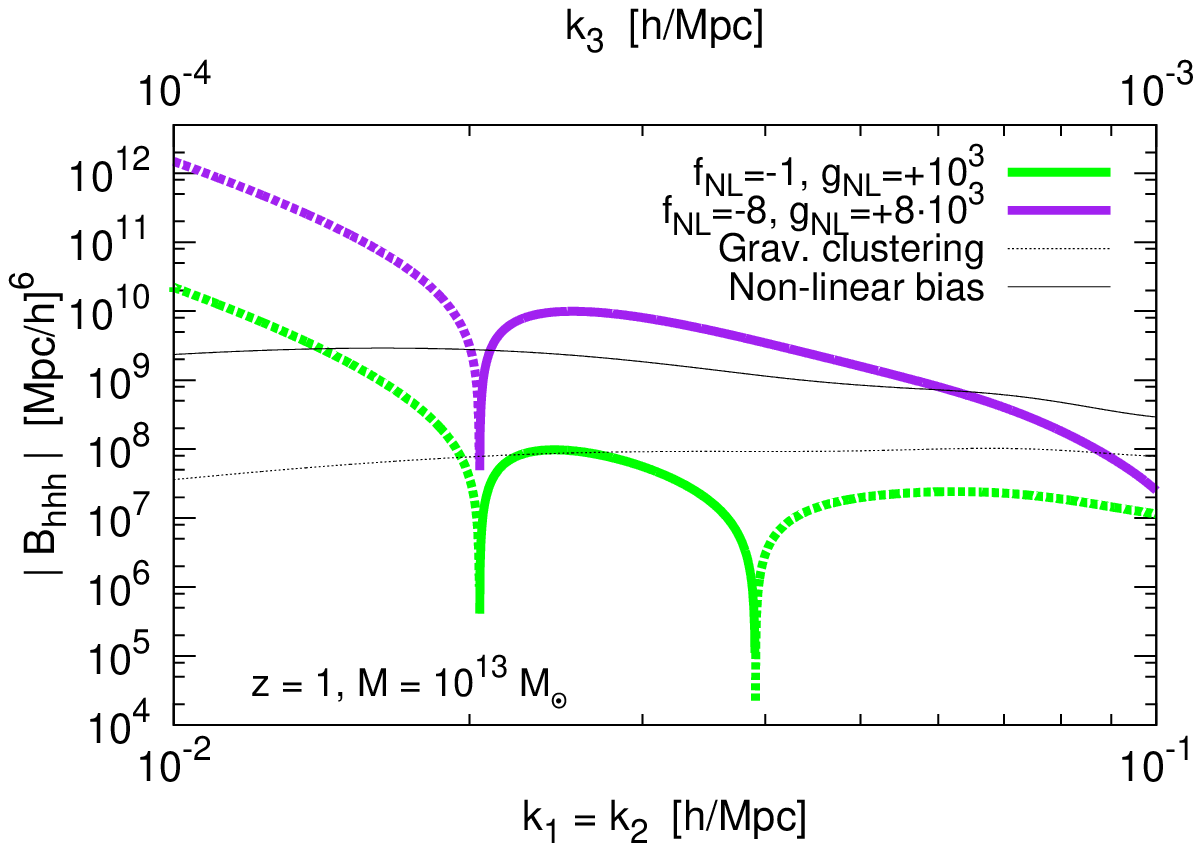}} 
\begin{caption}\\The plot shows the amplitude of halo bispectra as a function of scale. The colored lines
are bispectra associated with primordial nG only; the black lines are contributions to bispectra due to
non-linear gravitational clustering. The black dotted line corresponds to the gravitational clustering contributions to $B_{hhh}$ depending on the 
linear bias $b_{g\,0}$ only. The continuous line is the gravity bispectrum associated with non-linear bias $b_2$ (we take $b_2\,=\,2 b_{g\,0}$).
\end{caption}
\label{fig:six}
\end{figure}
 
 We notice that the amplitude of the bispectrum due to primordial nG tends to dominate at large scales
  (small $k$). However,  if the primordial nG parameters are chosen to be small, the bispectrum  induced by gravity
   is more important at smaller scales (see green line). For somewhat larger values of the nG parameters (with the
   magnitude of $\gnl$
    still below CMB constraints)
  the amplitude of the
   primordial 
bispectrum  
  can dominate over the gravity contributions
also at relatively small scales (see violet line). The  primordial bispectrum 
represented in the violet line
has non-trivial profile due to competing effects of primordial $\fnl$ and $\gnl$, in particular it has zeroes
  associated with the presence of $\gnl$: if such features can be detected in
  the bispectrum profile,
  they
    can be used  to help distinguishing among the effects of $\fnl$ and $\gnl$. 
  
  The plot above is only  indicative, in the sense that one should study in more detail the {\it combined}
  effects of non-linear gravity and primordial nG in shaping the bispectra, and not only comparing their
  relative amplitude. On the other hand,  it gives the idea that for interesting values of nG parameters
  the  primordial contribution can lead to a rich profile for  the bispectrum,   controlled by the primordial
  nG parameters, with an amplitude larger or comparable to the one due to gravitational clustering only.

\section{Discussion}\label{sec-disc}

Primordial nG characterizes   the interactions of the fields sourcing the density fluctuations that seed the large-scale structure of our universe.  In this work, 
 using a generalized univariate approach that we compared with other methods in the literature, 
we analytically and numerically investigated how the statistics 
 of LSS, in particular the study of bispectra of halos and matter, allow one to probe primordial nG.
  Our aim was to find
 ways to determine distinctive features of  the local non-Gaussian parameters $\fnl$ and $\gnl$, controlling
 respectively skewness and kurtosis of the probability distribution function, and 
 whose effects in the halo and matter power spectra are nearly degenerate.
 This investigation is important since although CMB constraints on   $\fnl$ have greatly improved with Planck, it is not clear 
 whether CMB measurements only can set stringent constraints on $\gnl$. 
 
 In the first part of this work, we showed analytically that the profiles of halo and matter bispectra have qualitative features
 that if measurable would clearly distinguish the effects of each non-Gaussian parameter.
  We exploited the scale dependence of the halo bias induced by primordial nG. 
  We worked in a simplified situation in which only linear bias and linear effects of gravitational clustering are taken into account, to single out the effects of primordial nG in the statistics
   of LSS. We have shown that the profile of the halo bispectrum as a function of the scale, for isosceles configurations of triangles in momentum space, has properties that provide simple, qualitative information on each non-Gaussian parameter. For example, the number and the position    of the zeros of this function depend on the size of $\gnl$,  and can be analytically  computed with our 
   formulae.
    If $\gnl$ is present, and its sign is opposite with respect to the sign of $\fnl$, the number of zeros of the bispectra increases
    as a function of the scale.
    %
   %
    Then, 
   we determined a particular combination of halo and matter bispectra and power spectra,
    denoted $C_{fnl}$, that  is proportional to $\fnl$ only and is 
   (nearly) independent on the value of $\gnl$.  A detection of this combination would be a clean probe of $\fnl$ with no contaminations associated with $\gnl$. 
 We then generalized our findings to the case in which multiple sources contribute to form the primordial density fluctuations, and
 studied how stochastic halo bias is affected by the presence of $\gnl$. We also provided new ways to study stochastic halo
 bias in terms of combinations of bispectra and power spectra.

 In the second part of this work, 
 we numerically confirmed and further developed our analytical results, focussing on a Press-Schechter approach to halo formation
 supplemented by appropriate non-Gaussian initial conditions. This analysis allowed us to apply our theoretical formulae for the bispectra focussing
 on scales
 and redshifts that might be probed by future surveys. We confirmed our theoretical considerations, showing plots representing
 how the qualitative profiles of halo bispectra in suitable configurations depend on non-Gaussian parameters, and might be used
 to set constraints on them. 
   We ended our discussion with a preliminary analysis of the contributions of non-linear gravitational 
  clustering,  indicating that non-linear gravity tends to contaminate the bispectrum profiles, but that the qualitative features
  we investigated survive for  sufficiently large primordial non-Gaussian parameters.
   

\bigskip

Our results can be extended in various directions. 
From the theoretical side,  the non-linear  effects associated with 
 features of the mass function, 
bias and  gravitational clustering (the non-linear transformation from Lagrangian to Eulerian coordinates), as well as loop effects, should be properly included
 in the context of perturbation theory, in order to understand whether and how primordial nG controls the qualitative features of bispectra also after taking these non-linearities  into account.
   Having a good control of gravitational effects 
  would also allow one to reliably study the bispectra for smaller values of the scale  and 
 for configurations that are less squeezed than the ones we considered here, and extend the 
 validity of  our 
 results  in these regimes.
   We plan to explore this interesting issue in a future publication. Another direction of investigation is a more systematic study of observational
 applications of our results, trying to quantify whether the methods we propose
 are sufficiently efficient to
   set 
 constraints on $\gnl$ that are more stringent than those available from the CMB. An important extension of our work
 is thus to forecast the signal-to-noise of galaxy bispectra as a function of scale, for future surveys such as EUCLID, and thereby 
 test whether the features we predict are observable. Such study will exploit our theoretical inputs to develop optimised methods to distinguish the effects of $\fnl$ and $\gnl$ using galaxy bispectra and thus quantify their size and sign independently. 
    %
%
 
 

\subsection*{Acknowledgments}
We thank Tommaso Giannantonio and Kazuya Koyama for useful discussions. 
GT is supported by an STFC Advanced Fellowship ST/H005498/1.
DW is supported by STFC grant ST/K00090X/1. GT and DW thank  the participants
  of  the workshop ``Constraining the early universe in the post Planck era", funded by the Royal Society,
  for useful feedback  on this work.

\begin{appendix}
\section{Comparison with previous works}\label{app:comparison}

In this appendix, we compare our results with the ones obtained in two previous papers that studied galaxy bispectra.

\smallskip

We start discussing a comparison  
 with the paper of Jeong and Komatsu \cite{Jeong:2009vd}. 
 Jeong and Komatsu  
 were the first to investigate the halo bispectrum up to the nG parameter $\gnl$
  including the scale dependence of bias.
   It is important to point out the 
   connection between their results and  our findings. 
     We address the issue in this Appendix. 

First of all, let us briefly sketch the formalism\footnote{Note that we replace their notation to make it consistent with ours.} used
in \cite{Jeong:2009vd}. They assume
   that the halo density contrast can be expressed as a local bias expansion of the matter overdensity in Eulerian coordinates, whose Fourier transform is

\be
\delta_h(\vec{k})=B_1\delta_M(\vec{k})+\frac{B_2}{2}\left[\int\frac{d^3q}{(2\pi)^3}\delta_M(\vec{k}-\vec{q})\delta_M(\vec{q})-\sigma_M^2\delta^{(3)}(\vec{k})\right]+\dots,
\ee   
where
\be
\delta_M(\vec{k})=W_M(k)\left[\delta^{(1)}(\vec{k})+\delta^{(2)}(\vec{k})+\delta^{D}(\vec{k})+\dots\right]
\ee
and $\delta^{(1)}(\vec{k})=\alpha(k)\Phi(\vec{k})$ , while $\delta^{(n)}(\vec{k})$ is the $n$-th order quantity of the linear density contrast, given by perturbation theory \cite{Bernardeau:2001qr}. The advantage of this approach is that non-linear evolution is automatically included when building the $3$-point function, even though calculations become harder.

On the other hand, the assumptions that led us to the halo density contrast of Eq.\eqref{coldh} are somewhat different. As we will show, the peak-background split method within a barrier crossing approach allows to re-obtain part of the results of \cite{Jeong:2009vd} in a simple way (at least in the squeezed limit) and extend them up to higher powers of nG parameters.
  However, since we have not yet accounted for the non-linear effects of
  gravity, a comparison is possible only between primordial nG effects at this stage. 
   Moreover, it is known  
   (see for example
     \cite{Matsubara:2011ck}) that the local Lagrangian  and  Eulerian bias are not compatible  in non-linear regimes.

By plugging Eqs.\eqref{eq:b1}, \eqref{eq:b2}, \eqref{eq:b3} into eq.\eqref{bshhh} and collecting all the terms proportional to the same nG parameter, we find:

 \bea
 B_{hhh}^{\fnl}&=&\fnl\Delta_0^2 b_g^3 \frac{\alpha^{(1)}\alpha^{(2)}\alpha^{(3)}}{k_1^3 k_2^3 k_3^3}(k_1^3+k_2^3+k_3^3)\simeq\,4\fnl\Delta_0^2 b_g^3\alpha^3_0\epsilon\,, \\
\nonumber \\ 
 B_{hhh}^{\fnl^2}&=&\fnl^2\Delta_0^2 b_g^2 \beta_2 \frac{\alpha^{(1)}\alpha^{(2)}\alpha^{(3)}}{k_1^3 k_2^3 k_3^3}\left[k^3_1\left(\frac{1}{\alpha^{(2)}}+\frac{1}{\alpha^{(3)}}\right)+\mbox{perm.}\right]\simeq\,4\fnl^2\Delta_0^2 b_g^2 \beta_2\alpha_0^2 \frac{\epsilon^3}{k^2}\,, \\
\nonumber \\ 
 B_{hhh}^{\gnl}&=&\gnl\Delta_0^2 b_g^2 \beta_2 \frac{\alpha^{(1)}\alpha^{(2)}\alpha^{(3)}}{k_1^3 k_2^3 k_3^3}\left(\frac{k_1^3}{\alpha^{(1)}}+\frac{k_2^3}{\alpha^{(2)}}+\frac{k_3^3}{\alpha^{(3)}}\right)\simeq\,6\gnl\Delta_0^2 b_g^2 \beta_2\alpha_0^2\frac{\epsilon}{k^2},
 \eea
where the approximated results are obtained by using $\alpha(k)\,=\,\alpha_0\,k^2$ and $T(k/\epsilon)\simeq T(k)$, in the
 squeezed configuration $k=k_1=k_2=\epsilon\,k_3$ with large
 $\epsilon$. The term $B_{hhh}^{\fnl}$ is the same primordial nG contribution found in \cite{Jeong:2009vd,Scoccimarro:2003wn,Sefusatti:2007ih}, while the approximated results for $B_{hhh}^{\fnl^2}$ and $B_{hhh}^{\gnl}$ match respectively those of eq.$(50)$ and eq.$(49)$ of \cite{Jeong:2009vd} on large scales. However, a subtle difference between our results and those of \cite{Jeong:2009vd} arise, due to the different initial assumptions. The matching is truly correct only if the equalities $B_1\equiv b_g$, $\beta_2\equiv 2B_2$ hold: it would be interesting to understand the physical meaning of this fact. 

Some interesting consideration follow from the ratios of the previous quantities:
 \bea
 \frac{B_{hhh}^{\fnl}}{B_{hhh}^{\fnl^2}}&=&\frac{1}{\fnl}\frac{b_g}{\beta_2}\frac{\alpha_0 k^2}{\epsilon^2}\,, \\
\nonumber \\ 
 \frac{B_{hhh}^{\fnl}}{B_{hhh}^{\gnl}}&=&\frac{2}{3}\frac{\fnl}{\gnl}\frac{b_g}{\beta_2}\alpha_0 k^2\,, \\
\nonumber \\ 
 \frac{B_{hhh}^{\fnl^2}}{B_{hhh}^{\gnl}}&=&\frac{2}{3}\frac{\fnl^2}{\gnl}\epsilon^2.
 \eea
It is clear from the first ratio that the $\fnl^2$ term can dominate over the $\fnl$ one in the squeezed limit, as pointed out in \cite{Jeong:2009vd}. Moreover, the bispectrum distinguishes between $B_{hhh}^{\gnl}$ and $B_{hhh}^{\fnl}$ (second ratio) but, surprisingly,
 the $\gnl$ term shows the same scaling as the $\fnl^2$ one.
 
At this point, all the primordial nG contributions found in \cite{Jeong:2009vd} have been recovered but new terms come out from eq.\eqref{bmsqhhh}. We collect them explicitly here:

\bea
B_{hhh}^{\fnl^3}&=&\,\fnl^3 \Delta_0^2 b_g \beta_2^2 \left(\frac{\alpha^{(1)}}{k_2^3 k_3^3} + \frac{\alpha^{(2)}}{k_1^3 k_3^3} + \frac{\alpha^{(3)}}{k_1^3 k_2^3} \right)\simeq\,4 \fnl^3 \Delta_0^2 b_g \beta_2^2 \alpha_0 \frac{\epsilon^3}{k^4}\,,\\
\nonumber \\ 
 B_{hhh}^{\fnl\gnl}&=&\fnl\gnl\Delta_0^2 b_g\frac{\alpha^{(1)}\alpha^{(2)}\alpha^{(3)}}{k_1^3 k_2^3 k_3^3}\left\{b_g\beta_3\left[k^3_1\left(\frac{1}{\alpha^{(2)}}+\frac{1}{\alpha^{(3)}}\right)+\mbox{perm.}\right]+\beta_2^2\left[\frac{k^3_1}{\alpha^{(1)}}\left(\frac{1}{\alpha^{(2)}}+\frac{1}{\alpha^{(3)}}\right)+\mbox{perm.}\right]\right\}\simeq
   \nonumber\\
 &\simeq& 2\fnl\gnl\Delta_0^2 b_g\epsilon^3\left(2\frac{b_g\beta_3\alpha_0^2}{k^2}+3\frac{\beta_2^2 \alpha_0}{k^4}\right)\,,\\
\nonumber \\ 
B_{hhh}^{\gnl^2}&=&\gnl^2\Delta_0^2 b_g\beta_2\beta_3\left(\frac{\alpha^{(1)}+\alpha^{(2)}}{k_1^3k_2^3}+\frac{\alpha^{(1)}+\alpha^{(3)}}{k_1^3k_3^3}+\frac{\alpha^{(2)}+\alpha^{(3)}}{k_2^3k_3^3}\right)\simeq 6\gnl^2\Delta_0^2 b_g\beta_2\beta_3\alpha_0\frac{\epsilon^3}{k^4}\,,\\
\nonumber \\ 
B_{hhh}^{\fnl^2\gnl}&=&\,\fnl^2\gnl\Delta_0^2\left[\beta_2^3\left(\frac{1}{k_1^3 k_2^3}+ \frac{1}{k_1^3 k_3^3} + \frac{1}{k_2^3 k_3^3}\right)+ 2\beta_2\beta_3 \left(\frac{\alpha^{(3)}}{k_1^3 k_2^3}+ \frac{\alpha^{(2)}}{k_1^3 k_3^3} + \frac{\alpha^{(1)}}{k_2^3 k_3^3}\right)\right]\simeq
\nonumber\\
&\simeq&\,2\fnl^2\gnl\Delta_0^2\epsilon^3\left(4\frac{b_g\beta_2\beta_3\alpha_0}{k^4} + 3\frac{\beta_2^3}{k^6} \right)\,,\\
\nonumber \\ 
B_{hhh}^{\fnl\gnl^2}&=&\,\fnl\gnl^2\Delta_0^2\left[2\beta_2^2\beta_3\left(\frac{1}{k_1^3 k_2^3}+ \frac{1}{k_1^3 k_3^3} + \frac{1}{k_2^3 k_3^3}\right)+b_g\beta_3^2\left(\frac{\alpha^{(3)}}{k_1^3 k_2^3}+ \frac{\alpha^{(2)}}{k_1^3 k_3^3} + \frac{\alpha^{(1)}}{k_2^3 k_3^3}\right)\right]
\nonumber\\
&\simeq&\,4\fnl\gnl^2\Delta_0^2\epsilon^3\left(\frac{b_g\beta_3^2\alpha_0}{k^4}+3\frac{\beta_2^2\beta_3}{k^6}\right)\,,\\
\nonumber \\ 
B_{hhh}^{\gnl^3}&=&\,\gnl^3 \Delta_0^2 \beta_2 \beta_3^2 \left(\frac{1}{k_1^3 k_2^3} + \frac{1}{k_1^3 k_3^3} + \frac{1}{k_2^3 k_3^3}\right)\simeq\,6\gnl^3 \Delta_0^2 \beta_2 \beta_3^2 \frac{\epsilon^3}{k^6}\,.
\eea

These additional terms appear naturally within our framework and suggest that higher powers of nG parameters can in principle dominate over the lower ones, extending what Jeong and Komatsu found for the ratio $B_{hhh}^{\fnl^2}/B_{hhh}^{\fnl}$. 
All the contributions above are important for the specific features of the halo bispectrum discussed in section \ref{sec:gnlLSS}.

\bigskip

Let us now compare our approach with the one by Baldauf et al in \cite{Baldauf:2010vn}. 
The easiest way to proceed is to reconsider our equation (\ref{coldh}): by substituting our
results for the bias coefficients $b_i$ (see eqs (\ref{eq:b1})-(\ref{eq:b3})), and using the relation $\delta_L
\,=\,\alpha \,\Phi_L$, we can schematically write our expression for  $\delta_h$
as an expansion
\be\label{bivarus}
\delta_h\,=\,b_g\,\delta_L+\beta_2\,\fnl\,\varphi_L
+\beta_3 \,\gnl\,\varphi_L+b_g\,\fnl\,\delta_L \varphi_L+\frac32\,\beta_2\,\gnl\,\varphi_L^2
+ b_g \,\gnl\,\delta_L\,\varphi_L^2\,,
\ee
where 
 to render the expression more
compact, we understand the subtraction of the averages  contained in (\ref{coldh}). This 
`bivariate' way of re-express our formula for the
halo bias makes clear the
connection with the bivariate approach of \cite{Giannantonio:2009ak,Baldauf:2010vn}, in which the Lagrangian halo bias  is
expanded in powers of $\delta_L$ and $\varphi_L$:
\be\label{bivar}
\delta_h\,=\,b_{10} \,\delta_L+b_{01} \,\varphi_L+\frac{b_{20}}{2} \,\delta_L^2
+\frac{b_{02}}{2} \,\varphi_L^2+b_{11}\,\varphi_L\,\delta_L+\,\dots\,,
\ee
where the bias coefficient $b_{ij}$ is defined as
\be
b_{ij}=\frac{1}{\bar{n}_{h}}\left(\frac{\partial^{i+j}n_h}{\partial^{i}\delta_L \partial^{j}\varphi_L}\right)\biggr|_{\delta_L=0, \varphi_L=0}\quad.
\ee
The values we find for the $b_{ij}$ coefficients are obtained by comparing eqs.
(\ref{bivarus}) with (\ref{bivar})\footnote{Note that in \cite{Baldauf:2010vn} the quantity $\nu$ corresponds to our $\nu^2$.}:
\bea
b_{10}&=&b_g
\\ \label{eqb01}
b_{01}&=&\beta_2 \fnl+\beta_3 \gnl
\\
b_{11}&=&b_g \fnl
\\
b_{02}&=&3 \beta_2 \,\gnl
\\
b_{20}&=&0
\eea
They
 differ from the ones of \cite{Giannantonio:2009ak,Baldauf:2010vn} for two reasons. The first reason is that, as
explained in the main text, in our approach we decided to perform a Taylor expansion at {\it linear order}
in $\delta$ and on all its correlation functions in Langrangian space.
In this way we   consider only   non-linear effects
induced by  primordial nG, and we exclude non-linearities associated with higher
derivatives of the mass function. This 
 approximation  is also justified by the fact that 
  higher
derivatives of the mass function along non-Gaussian parameters are under less reliable theoretical
control.  The second reason is that we include also
the effects of 
 (first) derivatives
of the mass function along $\fnl$ (that provide contributions proportional to $\beta_3$) that
are instead neglected in a bivariate approach. Hence, for example, 
the coefficient $b_{01}$ in eq (\ref{eqb01}), using our definition of $\beta_2$, $\beta_3$
 reads
\bea
b_{01}&=&\frac{2}{\bar{n}_h}\,\frac{\partial\, n_h}{\partial\,\ln{\sigma_M}}\,\fnl+\frac{3}{\bar{n}_h}\,\frac{\partial\, n_h}{\partial\,\fnl}\,\gnl
\eea
While the first term can be associated with the analogous result 
reported 
 in  \cite{Giannantonio:2009ak,Baldauf:2010vn}, the second term is absent in their case. Indeed, the bivariate approach produces the $\gnl/k^2$ dependence in the bias of the halo-halo and halo-matter power spectra as a contribution from the trispectrum \cite{Giannantonio:2009ak}. On the other hand, 
our expression for $b_{20}$ is vanishing because we decided  not to consider the second derivatives of $n_h$ along $\delta_L$.
 For the same reason of not considering such
 second derivatives, various contributions in the remaining coefficients $b_{ij}$ proportional to $\fnl^2$,
 found in \cite{Giannantonio:2009ak,Baldauf:2010vn} are absent in our formulae. 
 



\section{Inequalities among non-Gaussian parameters}\label{app:ine}
In this appendix we collect some technical results on inequalities among primordial non-Gaussian parameters, that
we use in Section \ref{sec:stochastic}. The most famous of these inequalities is dubbed Suyama-Yamaguchi 
inequality, and
relates
 the collapsed limit of the 4-pt function 
with the square of the squeezed limit of a 3-pt function
  of a given fluctuation $\delta$ in Fourier space
 %
 %
  \cite{Suyama:2007bg}
\be\label{nongi1}
\lim_{q\to0}\,\langle \delta_{k_1} \delta_{-k_1-q}  \delta_{k_2} \delta_{-k_2+q}  \rangle'\,\,
\ge \,\,\lim_{q\to0}  \frac{\langle \delta_{k_1} \delta_{-k_1-q}  \delta_{q}   \rangle'^2}{\langle \delta_{k_1}  \delta_{q}\rangle }
\ee
Using the definitions for $\taunl$ and $\fnl$ and specializing to the case 
of primordial curvature fluctuations, this inequality corresponds to  the Suyama-Yamaguchi inequality.
 It is 
 convenient to simplify the notation, expressing the squeezed limit of the 3-pt function in (\ref{nongi1}) in a synthetic way 
 as $\langle \delta [ \delta^2]\rangle$, 
  meaning that the momentum $q$ connecting $\delta $ and  $[ \delta^2]$ is sent to zero:
 this 
 notation emphasizes that the long mode $\delta_q$ with $q\to0$ modulates the 2-pt function $[ \delta^2]$.
 Analogously, the collapsed limit of the 4-pt function can be    expressed as $\langle  [ \delta^2]  [ \delta^2] \rangle$. With
 the help of this notation, inequality (\ref{nongi1}) succinctly  reads
 \be\label{nongi1b}
 \langle  [ \delta^2]  [ \delta^2] \rangle\,\ge\,\frac{\langle \delta  [ \delta^2] \rangle^2 }{\langle \delta \delta \rangle}
 \ee
 
In  \cite{Assassi:2012zq} a simple proof of this inequality has been provided, by inserting a complete set of normalized momentum eigenstates
$|n_{\vec k}\rangle \,\equiv\,|\delta_k\rangle/\langle \delta \delta \rangle^\frac12$ into the quantity  $ \langle  [ \delta^2]  [ \delta^2] \rangle$ of the left hand side of (\ref{nongi1b}).  The zero-momentum, long wavelength
eigenstate provides the square of the squeezed 3pt function in the right hand side, to which one has to add the
additional (positive definite) contributions that lead to the inequality above. In the collapsed limit and single source case, the
additional contributions vanish and one saturates the inequality.

Although less explored in the literature, it is also possible to build  further inequalities involving collapsed
and squeezed limits of higher point functions (see for example \cite{Suyama:2011qi,Byrnes:2011ri}):
 the virtue of the method 
 of \cite{Assassi:2012zq} is that the proofs of such inequalities 
  can be straightforwardly generalized to those cases.
As specific examples, we can write the following inequalities
satisfied by five and six point functions, in appropriate collapsed limits

\bea
\langle 
 [ \delta^3]  [ \delta^2]
\rangle
&\ge&
\frac{
\langle 
 [ \delta^3] \delta
\rangle
\langle 
 [ \delta^2] \delta \rangle
}{\langle \delta \delta \rangle} 
\label{nongi2}
\\\label{nongi3}
\langle 
 [ \delta^3]  [ \delta^3]
\rangle
&\ge&
\frac{
\langle 
 [ \delta^3] \delta
\rangle^2
}{\langle \delta \delta \rangle} 
\eea

Using the 
 methods of \cite{Assassi:2012zq},
it is relatively straightforward to 
 prove that the inequalities saturate to equalities in the single source case. 
  These inequalities have been used in Section \ref{sec:stocp}
   for physically  interpreting the notion of stochastic halo bias applied to power spectra of halos and matter. 
    When considering  stochastic halo bias  for the bispectra, Section \ref{sec:stocb} other inequalities are needed, that 
    involve subtler collapsed limits among higher order point functions. We collect them here
  \bea
{\langle
  [\delta^2]\, [\delta^2] \, [\delta^2]
  \rangle }
  &\ge&
  \frac{{\langle
  \delta\, [\delta^2] \, [\delta^2]
  \rangle }{\langle
  \delta\, [\delta^2]
  \rangle }}{\langle
  \delta \,  \delta
  \rangle }
 \\ 
   \langle
  \delta \,   [\delta^2]\, [\delta^2]  
  \rangle&\ge&
  \frac{
    \langle
  \delta \,\delta\,   [\delta^2]\rangle \langle\delta\, [\delta^2]  
  \rangle
  }{\langle
  \delta \,  \delta
  \rangle}
  \\ 
   \langle
  \delta \,\delta\,   [\delta^2]\rangle
   &\ge& \frac{
    \langle
  \delta \,\delta\,   \delta\rangle \langle\delta\, [\delta^2]  
  \rangle
  }{\langle
  \delta \,  \delta
  \rangle}
  \eea
  They can be analyzed and proved again with the methods of 
  \cite{Assassi:2012zq}.

\end{appendix}

\end{document}